\documentclass[authoryear]{elsarticle} 

\usepackage{fullpage}
\journal{Environmental Modelling and Software}

\usepackage{enumitem}

\usepackage{upgreek}
\usepackage{mathtools}
\usepackage{rotating}
\usepackage{amsmath}
\usepackage{amssymb}
\usepackage{bbm}
\usepackage{subcaption}
\usepackage{graphicx}
\usepackage{url}
\usepackage[linesnumbered]{algorithm2e}
\usepackage{placeins}

\newcommand {\un}[1]{\boldsymbol{#1}}

\newcommand {\PPC}{\texttt{covXtreme} }

\begin{document}

\begin{frontmatter}
	
	\title{\PPC : MATLAB software for non-stationary penalised piecewise constant marginal and conditional extreme value models\\(Second submission, March 2024)}
	
	\author[shelluk]{Ross Towe}
	\address[shelluk]{Shell Research Limited, London SE1 7NA, United Kingdom.}
	
	\author[shellnl]{Emma Ross}
	\address[shellnl]{Shell Global Solutions International BV, 1031 HW Amsterdam, The Netherlands.}	
	
	\author[shellnl]{David Randell}
	
	\author[shelluk,lancs]{Philip Jonathan}
	\address[lancs]{Department of Mathematics and Statistics, Lancaster University LA1 4YF, United Kingdom.}

	
	\begin{abstract}
    The \PPC software provides functionality for estimation of marginal and conditional extreme value models, non-stationary with respect to covariates, and environmental design contours. Generalised Pareto (GP) marginal models of peaks over threshold are estimated, using a piecewise-constant representation for the variation of GP threshold and scale parameters on the (potentially multidimensional) covariate domain of interest. 
	The conditional variation of one or more associated variates, given a large value of a single conditioning variate, is described using the conditional extremes model of Heffernan and Tawn (2004), the slope term of which is also assumed to vary in a piecewise constant manner with covariates. Optimal smoothness of marginal and conditional extreme value model parameters with respect to covariates is estimated using cross-validated roughness-penalised maximum likelihood estimation. Uncertainties in model parameter estimates due to marginal and conditional extreme value threshold choice, and sample size, are quantified using a bootstrap resampling scheme. Estimates of environmental contours using various schemes, including the direct sampling approach of Huseby et al. 2013, are calculated by simulation or numerical integration under fitted models. The software was developed in MATLAB for metocean applications, but is applicable generally to multivariate samples of peaks over threshold data. 
	The software and case study data can be downloaded from GitHub, with an accompanying user guide. 
	\end{abstract}
	
\end{frontmatter}

\section{Introduction} \label{Sct:Int}

Reasonable estimation of the characteristics of extreme environments is essential to quantify natural hazards, and assure the reliability and safe operability of man-made infrastructure. For example, the extreme ocean environment can be thought of as a multivariate random process in space and time, describing interacting wind, wave, current, tide and surge phenomena. Accurate theoretical modelling of \emph{extremes} from the ocean environment is problematic, because of the complexity of the numerical calculations involved. Practical risk assessment for marine and coastal structures and operations therefore typically requires the development of a statistical model from observational data for the joint behaviour of multiple meteorological-oceanographic (``metocean'') variables. 

Physical intuition and previous work has demonstrated that the marginal distribution of a variable such as ocean storm severity (quantified in terms of significant wave height, $H_S$) is dependent on covariates such as the direction in which the storm evolves, and the time of year in which the storm occurs (e.g. \citealt{RndEA15a}). Over the long term, there is evidence that storm severity also evolves slowly in time due to anthropogenic climate change as well as natural climate cycles (e.g. \citealt{EwnJnt23a}). It is important therefore that any statistical model for the extreme environment captures covariate non-stationarity. The nature and extent of covariate dependence varies across metocean variables. Experience also suggests that it is reasonable to assume that typical metocean variables are drawn from so-called ``max-stable'' distributions (e.g. \citealt{Cls01}). Therefore, typically conditional on covariates, the conditional distribution of threshold exceedances will be well approximated by the generalised Pareto (GP) distribution, provided that the threshold level is sufficiently high. This suggests that the marginal characteristics of metocean variables can be described reasonably using a non-stationary GP model (e.g. \citealt{DvsSmt90}, \citealt{ChvDvs05}, \citealt{RndEA15a}).

In addition we require that the statistical model also describes the \emph{joint} tail of all metocean variables in general. However the nature of extremal dependence between different metocean variables is generally unknown. The specification of the statistical model therefore needs to be sufficiently general to admit different extents of extremal dependence, the specifics of which are then estimated by fitting the model to data. The conditional extremes model of \cite{HffTwn04} is an attractive candidate, because it admits different classes of extremal dependence, it has a simple form and is relatively easily interpretable. There is also evidence that the nature and extent of extremal dependence also varies systematically with covariates (e.g. \citealt{JntEwnRnd13}, \citealt{RssEA17b}, \citealt{ShtEA20a}).

The sizes of data samples available (e.g. from direct observation of the environment) for modelling are typically small relative to the time-scales of the inference task. For example, in engineering design we typically need to characterise events which occur on average once in 1000 or 10000 years based on a sample of at most the order of 100 years of observation. In these circumstances, our estimate of the joint tail of the distribution of metocean variables is typically uncertain. It is critical that this uncertainty is captured and incorporated appropriately in our risk assessment. 

For individual metocean variables, the size of a rare occurrence is often quantified in terms of a \emph{return value} associated with some return period $T$ (=1000 years, say). The return value is defined straightforwardly as the quantile of the distribution of the annual maximum of the metocean variable with exceedance probability $1/T$, or (for large $T$, see \citealt{JntEA20}) the quantile of the distribution of the $T$-year maximum with non-exceedance probability $\exp(-1)$. In engineering design, the joint tail of the distribution of two (or sometimes three) metocean variables is often quantified in terms of an \emph{environmental design contour}; points on the contour are ``equally rare'' with respect to some criterion, related to the joint cumulative distribution function or density of the variables (e.g. \citealt{RssEA19}, \citealt{HslEA21}). Therefore estimates for the distribution of the annual maximum or that of the $T$-year maximum for individual metocean variables, and environmental contours for pairs or triplets of variables, are key outputs from the statistical analysis. 

There are many software tools available for applied extreme value analysis; \cite{StpGll06}, \cite{GllEA12}, \cite{Gll16} and \cite{BlzEA23} provide reviews. {In the statistical software literature, the R packages texmex (\citealt{texmex}), evgam (\citealt{evgam} and mgcv (\citealt{mgcv}) provide useful functionality for extreme value analysis.} The discussion of \cite{BlzEA23} notes the lack of good ``off the shelf'' software for practitioners, particularly incorporating more recent advances in methodology. Some software has been published for environmental contour estimation (e.g. \citealt{HslEA19}, \citealt{MckGll23}). The purpose of the \PPC software introduced here is to provide the environmental practitioner with a straightforward means to estimate marginal and joint tails of distributions of random variables, and quantify extremes in terms of return values and environmental design contours. \PPC accommodates covariate non-stationarity in both marginal and dependence behaviour, provides flexibility in estimating extremal dependence, and careful uncertainty quantification in inferences. 

The sophistication of the methodology underpinning \PPC has been set deliberately, based on the authors' experience of metocean applications, to accommodate necessary effects in a pragmatic manner, whilst avoiding unnecessary complexity. Extreme value analysis is in some senses a less mature research area (e.g. \citealt{DvsHsr15}) with numerous pressing challenges. For example, there are many approaches to modelling non-stationary marginal extremes; the work of \cite{Bll00}, \cite{Wod04},  \cite{BrzLng06}, \cite{HsrGnt16},  \cite{WodEA16}, \cite{WodEA17}, \cite{Yng19a}, \cite{Yng20}, \cite{ShaEA22} provide examples. With sufficiently rich data, complex extreme value models can be well estimated. However, in practice, typical samples tend only to support the adoption of relatively simple covariate models; for example, \cite{ZnnEA19a} show that a simple piecewise constant covariate representation (used by \PPC) provides marginal inferences competitive with those from more sophisticated P-spline (e.g. \citealt{ElrMrx10}) and Bayesian adaptive regression spline models (e.g. \citealt{Bll00}) for a metocean application.

Elements of the \PPC methodology, such as marginal and dependence modelling for extreme values, uncertainty quantification and environmental risk decision support, are reflected in recent developments in the environmental software and modelling literature. These typically address modelling of extremes of rainfall, river levels, temperatures and heatwaves, but sometimes less common environmental variables such as urban black carbon (\citealt{McjEA15}). For example, regarding marginal extreme value analysis, \cite{LcnNpl23} introduce the \texttt{EXTRASTAR} software for modelling of time-series of annual maxima, applied to rainfall. \cite{ZhEA22} use the generalised extreme value distribution to estimate return values of rainfall. \cite{DzSdlJ17} use extreme value analysis of block maxima within their \texttt{MENSEI-L} software. \cite{SssEA23} exploit extreme value analysis of block maxima in their model for river temperatures. \cite{CbsEA22} offer extreme value functionality within their \texttt{MarineTools.temporal} package. \cite{MngEA10} discuss non-stationary extreme value analysis of block maxima, using the Akaike Information Criterion (AIC) for optimal covariate parameterisation. \cite{ZhnEA22} consider non-stationary extreme value analysis of block maxima for heat wave tracking. Regarding joint modelling, \cite{ChnEA23} consider both marginal extreme value analysis and dependence modelling using copulas in their development of a drought index, and \cite{LiuEA21} consider vine copulas in their model for water level prediction. \cite{HaoEA17} develop the R package \texttt{drought} incorporating marginal and joint extreme value modelling using copulas, for applications to modelling and assessment. Regarding uncertainty quantification, \cite{VllEA23} present a stochastic model for future extreme temperatures, for decision support in infrastructure analysis, and the \texttt{ValueDecisions} software of \cite{HagEA22} considers uncertainty quantification for decision making, including some elements of extreme value analysis. More generally, \cite{CmpEA22} discuss extreme value analysis in the context of identifying outliers in environmental time-series, and \cite{Vrg16} offers extreme value functionality in their \texttt{DREAM} software for Bayesian inference using Markov chain Monte Carlo.

An early version of the \PPC software was published alongside \cite{RssEA17b} for modelling of storm surge, and \cite{RssEA19} for estimation of environmental contours. \cite{BorEA19} suggest that \PPC is useful for analysis of extreme ocean current profiles with depth. \cite{VnmEA20} use \PPC to compare different contour methods with response-based methods for extreme ship response analysis. \cite{GurEA23} have used their enhancement of an earlier version of \PPC for analysis of electrical signals in the human brain. The software has motivated the development of analogous prototype software PPL for marginal extreme value modelling with penalised piecewise-linear covariate representations (\citealt{BrlEA22}). The software has also been used as a pre-processor for transformation of data to standard marginal scale, allowing joint and conditional extreme value analysis (e.g. \citealt{ShtEA20a}; \citealt{ShtEA21}), and in metocean consultancy work.

\subsection*{Target audience for \PPC}
{\PPC was developed for oceanographic applications, but it can and has been used more widely. We believe that the reader will find \PPC potentially useful, if (a) there is interest in quantifying extremes using statistical analysis of a data set, and possibly estimation of extreme quantiles or ``return levels''; (b) the data are not stationary, and vary with known covariates; (c) there is interest in quantifying extremes of multiple variables at the same time; and (d) there is a need to quantify uncertainty carefully. Under these conditions, \PPC offers a pragmatic but statistically sound analysis.}

\subsection*{Objectives and layout of article}
The objective of the current article is to provide motivation and description of the covXtreme software, and illustrations of its use in the development of design conditions for ocean engineering. The layout of the article is as follows. Section 2 provides an overview of the software and the statistical methodology on which it is based. Sections 3 and 4 present case studies, involving a bivariate response and single covariate (Section 3), and trivariate response with 2-D covariate (Section 4). An accompanying user guide (available at \citealt{TowEA23a}) provides a detailed step-by-step description of developing a covXtreme model for ocean engineering data sets provided with the software.

\section{Overview of software and statistical methodology}  \label{Sct:OvrSft}
\PPC provides simple software for multivariate non-stationary extreme value analysis of peaks over threshold. The modelling framework of \PPC (a) is statistically straightforward to understand and use without sacrificing rigour, (b) is computationally efficient, (c) provides good estimation of key quantities of interest to the extreme value practitioner, and (d) provides realistic quantification of modelling uncertainties.

Extreme value analysis of peaks over threshold of metocean variables requires the satisfactory completion of a number of analysis steps. Inference typically involves using a sample of time-series data for a number of variables (corresponding to some period $P$ years of observation) as the basis for estimating the characteristics of extreme environments. These might include return values, associated values (e.g. \cite{TowEA23}) and design contours (e.g. \citealt{HslEA21}) corresponding to a return period of $T$ years, $T \gg P$). For reasonable extreme value inference of a single variable (such as significant wave height), key features of the data must be accommodated carefully in the analysis, including the serial correlation of time-series, the dependence of values of metocean variables on covariates such as direction and season. Appropriate model forms for tails of distributions should also be adopted. For peaks over threshold analysis of a single variable, choice of threshold level can be source of considerable uncertainty. Moreover, different metocean variables (such as wind speed and significant wave height) are likely to be intercorrelated, and the nature of this dependence must be characterised carefully especially for extremes of one or more variables, using appropriate model forms; in these models, choice of threshold is again not always straightforward. Further, since samples of peaks over threshold are generally small (e.g. for high thresholds), it is critical to quantify uncertainty in return values and associated extremal quantities thoroughly using well-understood estimators. 

\begin{figure}[h!]
	\centering
	\includegraphics[width=0.75\textwidth]{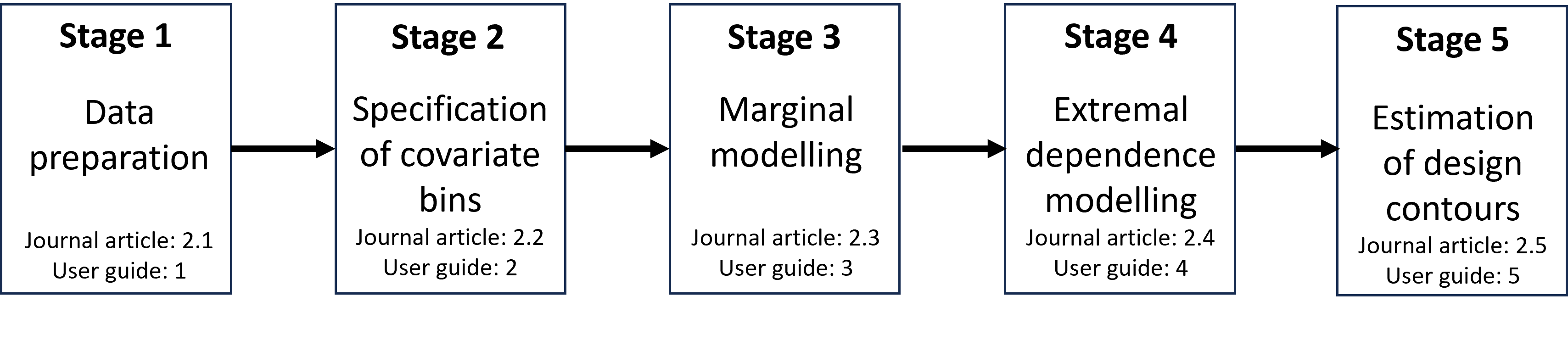}
	\caption{{Schematic of \PPC workflow. The five analysis stages are executed in sequence. Generally, at each stage, model tuning based on refinement of modelling hyper-parameters is necessary, guided by diagnostic information, before the next stage is attempted. Section numbers in the main article and user guide, providing relevant details for each stage of analysis, are given.}}
	\label{Fgr:covXtremeFlowScheme}
\end{figure}

The structure of the \texttt{covXtreme} software is written to reflect the sequential nature of practical extreme value analysis, as a set of MATLAB functions (see Section~\ref{Sct:OvrSft:PrfTpcAnl}), illustrated in the workflow of Figure~\ref{Fgr:covXtremeFlowScheme}. 
For a given application, the five stages are executed sequentially, leading the user through the inference, addressing important analysis considerations in order. A typical analysis stage involves the specification of tuning parameters, the appropriate choice of which is informed by diagnostic information generated at that stage; generally it will therefore be necessary to repeat the analysis stage until satisfactory diagnostics are achieved, before moving on to the next analysis stage. 

Given a sample of multivariate time-series of metocean variables, a typical analysis would proceed as described below. Note, for ease of reference, that subsection numbering  \ref{Sct:OvrSft:DatPrp}-\ref{Sct:OvrSft:EstDsgCnt} here corresponds to the numbering of stages 1-5 of the analysis procedure, and also to the numbering of sections in the \PPC user guide.

\subsection{Data preparation} \label{Sct:OvrSft:DatPrp}
The main aspect of data preparation is the isolation of storm peak events. Storm peak events correspond to peaks over threshold for a dominant variable (e.g. significant wave height for a wave-dominated metocean application), \emph{associated} values of all other variables of interest (e.g. wind speed) per event, and storm peak covariate values for all relevant covariates (e.g. direction and season). Storm peak events are isolated using the procedure described by \cite{EwnJnt08}, and discussed in Section 1 of the covXtreme user guide. Storm peak events are then assumed to be conditionally independent given (storm peak) covariates.

Mathematically, consider a sample $\ddot{Y}_1(t_i), \ddot{Y}_2(t_i), ..., \ddot{Y}_D(t_i)$ of multivariate time-series for $D$ metocean variables observed at regularly-spaced time points $t_i$, $i=1,2,3, ...$  over some period. Data preparation requires the isolation of a sample {of values} $\{\dot{y}_{i1}, \dot{y}_{i2}, ..., \dot{y}_{iD}\}_{i=1}^N$ of $N$ storm peak events (by convention for the first variable) and associated events (for the remaining variables) summarising the peak characteristics of each of $N$ storms observed, and corresponding storm peak covariate values $\{x_{i1}, x_{i2}, ..., x_{iC}\}_{i=1}^N$ for $C$ storm peak covariates $X_1, X_2, ..., X_C$ defined on some domain $\mathcal{X}$. {(Note that the ``double dot'' notation (e.g. $\ddot{Y}$) indicates \emph{time-series} variables from which \emph{storm peak} events, indicated by ``single dot'' notation (e.g. $\dot{Y}$) must be isolated.)} {Storm peak events are identified as local maxima (between successive up- and down-crossings of a given level) of $\ddot{Y}_1$. Associated values for a storm are the values of the other random variables at the time of occurrence of the storm peak event.} Storm peak and associated values are taken to be conditionally-independent given covariates, in the sense that $\dot{y}_{id}$ can be viewed as an independent draw from $\dot{Y}_d|(X_1=x_{i1}, X_2=x_{i2}, ..., X_C=x_{iC})$, for $i=1,2,...,N$, $d=1,2,...,D$.   

Note that \PPC also provides functionality to simulate data with known characteristics for checking of the performance of the statistical methodology.

\subsection{Specification of covariate bins} \label{Sct:OvrSft:SpcCvrBns}
\PPC adopts a piecewise-constant parameterisation for the distribution of peaks over threshold on a partition of the covariate domain comprised of elements referred to as ``covariate bins''. Covariate bins are not estimated as part of the analysis, but must be specified carefully by the user prior to the analysis, for reasonable inference.  Briefly, the user is required to partition the domain of each covariate (independently) into a set of intervals. The resulting covariate bins over all covariates are then simply Cartesian products of these intervals, {and hence a rectangular grid of covariate bins}. \PPC assumes, \emph{within} a given covariate bin, that the statistical properties of a storm peak or associated variable are no longer dependent on covariates. Diagnostic plots illustrating features such as the variation of storm peak and associated variables with individual covariates, are provided to inform reasonable choice of marginal partitions. Illustrations of the specification of covariate bins are given in Section 2 of the \PPC user guide. 

Mathematically, each observation $\dot{y}_{i1}, \dot{y}_{i2}, ..., \dot{y}_{iD}$  is allocated to one of $B$ covariate bins by means of an allocation vector $A$, with $A(i)=b$ mapping observation (with index) $i$ to bin (with index $\mathcal{B}=$) $b$, $i=1,2,...,N$, $b=1,2,...,B$. All observations within a specified covariate bin are assumed to have common extreme value characteristics, specified in terms of the parameters of the marginal model for peaks over threshold from that bin for each of the $D$ components of the observation. Hence in particular, all threshold exceedances of storm peak variable $\dot{Y}_d$, $d=1,2,...,D$ from covariate bin $b$ can be viewed as independent draws from a generalised Pareto distribution with common shape, scale and threshold parameter values.

\subsection{Marginal modelling} \label{Sct:OvrSft:MrgMdl}
A two-stage modelling procedure is used to describe the marginal distribution of storm peak and associated variables. First a three-parameter gamma distribution is fitted independently to all the data for each covariate bin in turn, providing a good description of the bulk of the distribution of storm peak and associated variables on the covariate domain. The extreme value threshold for each covariate bin is then set to the quantile of the corresponding gamma distribution with pre-specified non-exceedance probability. Finally the distribution of threshold exceedances per covariate bin is estimated using a non-stationary generalised Pareto distribution. Since generalised Pareto shape parameter is typically more difficult to estimate than scale, its value is assumed fixed but unknown across all covariate bins. Moreover, variation of the generalised Pareto scale parameter on the covariate domain is regulated using roughness penalisation, set to maximise the predictive performance of the generalised Pareto model on a hold-out sample within a cross-validation scheme. We judge from prior experience that these assumptions regarding the variation of generalised Pareto parameters are reasonable, and confirm using diagnostics during analysis that the assumptions give reasonably well-fitting models. By reducing the number of degrees of freedom for model fitting, the assumptions are a possible source of estimation bias, but also of a reduction in variance in estimated model parameters and return values. Further details and illustrations of marginal modelling are given in Section 3 of the \PPC user guide. 

Mathematically, for each variable $\dot{Y}_d$, $d=1,2,...,D$, and covariate bin $\mathcal{B}=b$, $b=1,2,...,B$ independently, we estimate a three-parameter gamma distribution using maximum likelihood estimation with sample likelihood
\begin{eqnarray}
	\mathcal{L_{\text{Gmm}}}\left( \omega_{bd}, \kappa_{bd}, l_{bd}; \{\dot{y}_{id}\}_{i=1}^N \right) = 	\prod_{i;A(i)=b}  f_{\text{Gmm}}(\dot{y}_{id}; \omega_{bd}, \kappa_{bd}, l_{bd})
\end{eqnarray}
for sample $\{\dot{y}_{id}\}_{i=1}^N$, where $f_{\text{Gmm}}$ is the density of the gamma distribution with shape $\omega_{bd} \in \mathbb{R}$, scale $\kappa_{bd}>0$ and location $l_{bd} \in \mathbb{R}$ given by 
\begin{eqnarray}
	f_{\text{Gmm}} \left(\dot{y}_{id}; \omega_{bd}, \kappa_{bd}, l_{bd} \right) = 
	\left( \kappa_{bd}^{-\omega_{bd}}/\Gamma(\omega_{bd})\right) (\dot{y}_{id} - l_{bd})^{\omega_{bd} - 1} \exp \left( (\dot{y}_{id}-l_{bd})/\omega_{bd} \right)
	\label{eq:DnsGmm}
\end{eqnarray}
where $\Gamma(\bullet)$ is the gamma function. In practice for fitting the gamma model, the location parameters $\{l_{bd}\}$ are first estimated using a low empirical quantile of the sample per variate per covariate bin, and the remaining shape and scale parameters estimated using maximum likelihood. Next we calculate the extreme value threshold as $\psi_{bd} = F_{\text{Gmm}}^{-1}(\tau_d; \hat{\omega}_{bd}, \hat{\kappa}_{bd}, \hat{l}_{bd})$ using the estimated gamma parameters, for the pre-specified non-exceedance probability $\tau_d$, where $F_{\text{Gmm}}$ is the cumulative distribution function of the gamma distribution. The marginal sample likelihood for threshold exceedances of $\psi_{bd}$ for variable  $\dot{Y}_d$ over all covariate bins can therefore be written
\begin{eqnarray}
	\mathcal{L}_{\text{GP}}\left(\xi_d, \{\nu_{bd}\}_{b=1}^B; \{\dot{y}_{id}\}_{i=1}^{N}, \{\psi_{bd}\}_{b=1}^B \right) = \prod_{b=1}^{B}  \prod_{i;A(i)=b;\dot{y}_{id}>\psi_{bd}(\tau_d)} f_{\text{GP}}(\dot{y}_{id}; \xi_d, \nu_{bd}, \psi_{bd})
\end{eqnarray}
where $f_{\text{GP}}$ is the density of the generalised Pareto distribution with shape $\xi_{d} \in \mathbb{R}$ and scale $\nu_{bd}>0$ given by
\begin{eqnarray}
	f_{\text{GP}}(\dot{y}_{id}; \xi_d, \nu_{bd}, \psi_{bd}) = (1/\nu_{bd}) [1 + (\xi_d/\nu_{bd}) \left(\dot{y}_{id}-\psi_{bd}\right) ]_+^{-1/\xi_d-1}
	\label{eq:DnsGP}
\end{eqnarray}
where $[A]_+=A$ when $A>0$, and $=0$ otherwise. The corresponding cumulative distribution functions of the gamma and generalised Pareto distributions are
\begin{eqnarray}
	F_{\text{Gmm}}(\dot{y}_{id}; \omega_{bd}, \kappa_{bd}, l_{bd}) = 
	(1/\Gamma(\alpha)) \quad \gamma(\omega_{bd}, \kappa_{bd} (\dot{y}_{id}-l_{db}))
\end{eqnarray}
where $\gamma(\bullet, \bullet)$ is the lower incomplete gamma function, and 
\begin{eqnarray}
	F_{\text{GP}}(\dot{y}_{id}; \xi_{d}, \nu_{bd}, \psi_{bd}) = 
	1- \left[1 + (\xi_d/\nu_{bd}) \left( \dot{y}_{id}-\psi_{bd} \right) \right]_+^{-1/\xi_d}  .
\end{eqnarray}

\textbf{Optimal predictive performance} for the marginal generalised Pareto model is achieved using roughness-penalisation for the scale parameters $\{\nu_{bd}\}_{b=1}^B$ across covariate bins. The corresponding penalised negative log likelihood takes the form 
\begin{eqnarray}
	&&-\log \mathcal{L}_{\text{GP}}^*(\xi_d, \{\nu_{bd}\}_{b=1}^B; \{\dot{y}_{id}\}_{i=1}^{N}, \{\psi_{bd}\}_{b=1}^B) \nonumber \\  &=& - \log \mathcal{L}_{\text{GP}}(\xi_d, \{\nu_{bd}\}_{b=1}^B; \{\dot{y}_{id}\}_{i=1}^{N}, \{\psi_{bd}\}_{b=1}^B) + \lambda_d \left( \frac{1}{B} \sum_{b=1}^B \nu_{bd}^2 - \left[ \frac{1}{B} \sum_{b=1}^B  \nu_{bd} \right]^2 \right).
	\label{eq:LamMM}
\end{eqnarray}
The smoothness penalty $\lambda_d$ controls the extent to which the generalised Pareto scale varies across covariate bins, {and the penalty is proportional to the variance of $\nu$ over covariate bins}. Parameters can be estimated to minimise the penalised negative log likelihood for each variable $\dot{Y}_d$, $d=1,2,...,D$  in turn for each of a set of pre-specified values for $\lambda_d$. The optimal value $\lambda_d^\circ$ of $\lambda_d$ is chosen to maximise predictive likelihood for a hold-out sample within a $k$-fold cross-validation procedure. Gamma parameter estimates per covariate bin, and generalised Pareto parameter estimates evaluated using the full sample for $\lambda_d=\lambda_d^\circ$ are carried forward to subsequent inference.

Since the numbers of parameters in the various marginal models above are relatively small, a simplex search procedure provides a straightforward approach to parameter estimation by minimisation of (penalised) negative log likelihoods. 

\textbf{Threshold selection} for extreme value analysis of peaks over threshold is an important consideration (e.g. \citealt{NrtJnt11},  \citealt{ScrMcd12}, \citealt{Wds16}). Within a \PPC analysis, multiple marginal models based on different random choices of threshold non-exceedance probabilities $\tau_d \in \mathcal{I}_{\tau_d} \subseteq [0,1)$ are constructed, $d=1,2,...,D$. The user's task is to specify the interval $\mathcal{I}_{\tau_d}$ for each variable $\dot{Y}_d$. As explained in Section~3 of the user guide, this choice is aided by numerous diagnostic plots, including examination of the stability of the estimated value of $\xi_d$ as a function of $\tau_d$.

\textbf{Uncertainty quantification} for marginal inference is performed using a non-parametric bootstrap procedure. The original sample of storm peak and associated values is resampled with replacement. For each variable $\dot{Y}_d$, $d=1,2,...,D$, the full marginal extreme value analysis is then repeated using the bootstrap resample together with a new random selection of $\tau_d$. The outcome of the complete marginal inference is then quantified in terms of sets of parameter estimates ( $\{ \xi_d^r, \{\nu_{bd}^r, \psi_{bd}^r, \omega_{bd}^r, \kappa_{bd}^r, l_{bd}^r\}_{b=1}^B; \tau_d^r, \lambda_d^\circ \}_{d=1}^D$) for each of $R$ (typically $=100$ or $250$) bootstrap resamples (where superscript $r$ indicates an estimate from a resample). Typically, the value of $\lambda_d^\circ$ estimated using the original sample is adopted for all bootstrap resamples also, although the option to estimate a new optimal roughness coefficient per bootstrap resample is provided.

The \textbf{distribution of the $T$-year maximum event} per covariate bin, and over all covariate bins can then be estimated (or sampled) under the fitted marginal model. From above, the full marginal model for storm peak or associated variate $\dot{Y}_d$ in covariate bin $\mathcal{B}=b$ is 
\begin{equation}
	F_{\text{GmmGP}}(y; \omega_{bd}, \kappa_{bd}, l_{bd}, \xi_{d}, \nu_{bd}, \psi_{bd}) =
	\begin{cases}
		F_{\text{Gmm}}(y; \omega_{bd}, \kappa_{bd}, l_{bd}) &\text{ for } A(i)=b, \dot{y}_{id} \le \psi_{bd}, \\
		\tau_d + (1-\tau_d) F_{\text{GP}}(y; \xi_{d}, \nu_{bd}, \psi_{bd}) &\text{ for } A(i)=b, y > \psi_{bd}.	 	 
	\end{cases}
	\label{eq:MrgFll}
\end{equation}

Then under the model, the distribution of a random occurrence of $\dot{Y}_d$ from any covariate bin is
\begin{eqnarray}
	F_{\dot{Y}_d}(y; \{ \omega_{bd}, \kappa_{bd}, l_{bd}, \nu_{bd}, \psi_{bd} \}_{b=1}^B, \xi_{d} ) = \sum_{b=1}^B p_{b} F_{\text{GmmGP}}(y; \omega_{bd}, \kappa_{bd}, l_{bd}, \xi_{d}, \nu_{bd}, \psi_{bd})
	\label{eq:MrgCdfRndEvn}
\end{eqnarray}
where $p_{b}$ is an empirical estimate of the probability of observing a storm event in covariate bin $b$. If we further assume that the number $N$ of storms in $T$-years is Poisson-distributed with mean $T \rho$, where $\rho$ is an empirical estimate for the number of storms per annum, and suppressing parameter dependence for brevity, the distribution of the $T$-year maximum is simply
\begin{eqnarray}
	F_{\dot{Y}_d T\text{-year}}(y) &=& \sum_{k=0}^\infty \mathbb{P}(N=k) F_{\dot{Y}_d}^k(y) 
	= \sum_{k=0}^\infty \left( \exp(-T \rho) (T\rho)^k / k! \right) F_{\dot{Y}_d}^k(y) \nonumber \\
	&=& \exp\left( -T \rho \ (1-F_{\dot{Y}_d}(y))\right) .
	\label{eq:MrgTYer}
\end{eqnarray}
We can use a similar approach to estimate and sample from the distribution of the $T$-year maximum of $\dot{Y}_d$ for any combination of covariate bins, by restricting the set of covariate bins over which the summation is performed in Equation~\ref{eq:MrgCdfRndEvn} (and linearly scaling the values of $\{p_b\}$ so that they sum to unity).

\subsection{Extremal dependence modelling} \label{Sct:OvrSft:ExtDpnMdl}
Given non-stationary marginal models for storm peak and associated variables, we next seek to describe the nature of the dependence between them for large values of the storm peak variable. This is achieved using the conditional extremes model of \cite{HffTwn04}. Under the fitted conditional extremes model, we can then estimate the characteristics of the joint distribution of all associated variables given a large storm peak, and thereby estimate joint environmental design conditions and design contours. The conditional extremes model is specified for sets of variables on a common standard Laplace marginal scale, rather than their original physical scales. For this reason, a necessary first step is to transform the sample of storm peak and associated values to this scale, using the fitted marginal models. Incorporation of covariate effects is generally important for extremal dependence modelling, and these can be accommodated using conditional extremes models of different complexities with respect to covariates. The \PPC software allows any number of the parameters of the conditional extremes model to vary with covariates. Nevertheless, experience suggests that the data indicate the need for covariate non-stationarity of just one model parameter (the ``slope'' parameter, see below) for typical metocean applications. We therefore describe this specific model form as a recommended ``default'' approach here. Further details and illustrations of extremal dependence modelling are given in Section 4 of the \PPC user guide. 

Mathematically, \PPC provides a model for the joint conditional tail $(\dot{Y}_2, \dot{Y}_3 \dots \dot{Y}_D | \dot{Y}_1=\dot{y})$ for large $\dot{y}$, using an extension of the conditional extremes model of \cite{HffTwn04} on standard Laplace (and optionally, as originally, on standard Gumbel)  scale. Inference therefore requires that we transform the storm peak and associated value sample $\{\dot{y}_{i1}, \dot{y}_{i2}, ..., \dot{y}_{iD}\}_{i=1}^N$ of variables $\{\dot{Y}_{i}, \dot{Y}_{2}, ..., \dot{Y}_{D}\}$ to the corresponding Laplace scale sample $\{y_{i1},y_{i2}, ..., y_{iD}\}_{i=1}^N$ for variables $\{Y_1, Y_2, ..., Y_D\}$. 

\subsubsection{Marginal transformation to standard Laplace scale}
The marginal transformation to standard Laplace scale is achieved using the probability integral transform such that for $i=1,2,...,N$, $d=1,2,...,D$, $b=1,2,...,B$
\begin{eqnarray}
	F_{\text{GmmGP}}(\dot{y}_{id}; \omega_{bd}, \kappa_{bd}, l_{bd}, \xi_{d}, \nu_{bd}, \psi_{bd}) = F_{\text{Lpl}}(y_{id}) \text{ for } A(i)=b
\end{eqnarray}
where $F_{\text{GmmGP}}$ is the marginal cumulative distribution function of storm peak variable $\dot{Y}_d$, for sets of parameters of marginal gamma and generalised Pareto models from Equation~\ref{eq:MrgFll}, and $F_{\text{Lpl}}$ is the cumulative distribution function of the standard Laplace distribution given by $F_{\text{Lpl}}(y) = 0.5 \exp(-|y|)$ for $y \le 0$ and $= 1 - 0.5 \exp(-|y|)$ otherwise.

\subsubsection{Conditional extremes modelling}
The Laplace-scale sample $\{y_{i1}, y_{i2}, ..., y_{iD}\}_{i=1}^N$ from random variables $\{Y_1, Y_2, ..., Y_D\}$ is next characterised using the conditional extremes model, for values $y$ of the conditioning variate $Y_1$ above a dependence threshold $\phi(\tilde{\tau})=F^{-1}_{Lpl}(\tilde{\tau})$, for which the conditional extremes model is assumed to hold, for carefully specified non-exceedance probability $\tilde{\tau}$. For $y>\phi(\tilde{\tau})$, in covariate bin indexed $\mathcal{B}=b$, the recommended non-stationary model takes the form
\begin{eqnarray}
	(Y_2, Y_3 \dots Y_D)|(Y_1 = y, \mathcal{B}=b) = (\alpha_{b2}, \alpha_{b3},\ldots,\alpha_{bD}) y + y^{(\beta_{2},\beta_{3},\ldots,\beta_{D})} \boldsymbol{Z}
	\label{eq:HTMdl}
\end{eqnarray}
with linear slope parameters $\alpha_{bd'} \in [-1, 1]$, $d'=2,3,...,D$ varying across covariate bins, scalar exponent parameters $\beta_{d'} \in [-\infty, 1]$ common to all covariate bins, and residual random variable $\boldsymbol{Z} = (Z_2, Z_3, ..., Z_D)\in \mathbb{R}^{D-1}$ whose distribution is unknown. 

For estimation of slope and exponent parameters {only}, we assume that each component $Z_{d'}$ of $\boldsymbol{Z}$ is independently distributed according to
\begin{eqnarray}
	Z_{d'} = \mu_{d'} + \sigma_{d'} W_{d'}
	\label{eq:HTRsd}
\end{eqnarray}
for mean $\mu_{d'} \in \mathbb{R}$, scale $\sigma_{d'}>0$ and random variable $W_{d'} \in \mathbb{R}$ following a generalised Gaussian (or delta-Laplace) distribution with zero mean, unit variance and shape $\delta_{d'}$. {$\mu_{d'}$ and scale $\sigma_{d'}$ are treated as nuisance parameters in the estimation, and are not needed for subsequent inferences under the fitted model.} For general mean $m$ and variance $s^2$, the corresponding density $f_{\text{GG}}$ of the generalised Gaussian distribution is
\begin{eqnarray}
	f_{\text{GG}}(w;m,s^2,\delta_{d'})=\frac{\delta_{d'}}{2\kappa(\delta_{d'}) s \Gamma(1/\delta_{d'})} \exp\left\{-\left\lvert\frac{w-m}{\kappa(\delta_{d'}) s}\right\rvert^{\delta_{d'}}\right\},
\end{eqnarray}
where $\kappa(\delta_{d'})^{2}=\Gamma(1/\delta_{d'})/\Gamma(3/\delta_{d'})$. The value of exponent $\delta_{d'} \in \{1,2\}$ is user-specified in \PPC. $\delta_{d'}=1$ imposes a Laplace distribution (with zero mean and unit variance) on $W_{d'}$, appropriate when the dependence between $Y_{d'}$ and $Y_1$ is low. $\delta_{d'}=2$ imposes a standard Gaussian distribution on $W_{d'}$, the original assumption of \cite{HffTwn04}, appropriate otherwise. For estimation purposes therefore, from the properties of the generalised Gaussian distribution, $Y_{d'}|(Y_1=y)$ is assumed to follow a generalised Gaussian distribution with mean $m_{bd'}=\alpha_{bd'} y + \mu_{d'} y^{\beta_{d'}}$ and standard deviation $\zeta_{d'}=y^{\beta_{d'}} \sigma_{d'}$.  

The conditional dependence likelihood $\tilde{\mathcal{L}}_{d'}$ for the model $Y_{d'}|(Y_1=y)$ and $y>\phi(\tilde{\tau})$ can then be written
\begin{eqnarray}
	\tilde{\mathcal{L}}_{d'} \left(\{\alpha_{bd'}\}_{b=1}^B, \beta_d, \mu_{d'}, \sigma_{d'}; \{y_{id}\}_{i=1,d=1}^{N,D}, \delta_{d'}, \tilde{\tau} \right) = 
	\prod^B_{b=1}\prod_{\stackrel{i : b=A(i)}{y_{i1}>\phi(\tilde{\tau})}} f_{\text{GG}}\left(y_{id};\alpha_{bd'} y_{i1} + \mu_{d'} y_{i1}^{\beta_{d'}},(y_{i1}^{\beta_{d'}} \sigma_{d'})^2,\delta_{d'}\right).
	\label{eq:HTLkl}
\end{eqnarray}
As for marginal models, we regulate the smoothness of $\{\alpha_{bd'}\}_{b=1}^B$ optimally on the covariate domain using a cross-validation procedure, selecting a value $\tilde{\lambda}_{d'}^\circ$ for roughness coefficient $\tilde{\lambda}_{d'}$ in roughness-penalised negative log likelihood
\begin{eqnarray}
	&&-\log \tilde{\mathcal{L}}_{d'}^*\left(\{\alpha_{bd'}\}_{b=1}^B, \beta_{d'}, \mu_{d'}, \sigma_{d'}; \{y_{id}\}_{i=1,d=1}^{N,D}, \delta_{d'}, \tilde{\tau} \right) \nonumber \\ 
	&=& 
	-\log \tilde{\mathcal{L}}_{d'} \left(\{\alpha_{bd'}\}_{b=1}^B, \beta_{d'}, \mu_{d'}, \sigma_{d'}; \{y_{id}\}_{i=1,d=1}^{N,D}, \delta_{d'}, \tilde{\tau} \right)
	+ \tilde{\lambda}_{d'} \left( \frac{1}{B} \sum_{b=1}^B \alpha_{bd'}^2 - \left[ \frac{1}{B} \sum_{b=1}^B  \alpha_{bd'} \right]^2 \right).	
	\label{eq:HTPNLL}
\end{eqnarray}
to maximise predictive performance on cross-validation hold-out samples. Further, it is important to confirm reasonable choice of the dependence threshold $\tilde{\tau}$ by inspection of diagnostic plots, for example of stability of estimated parameters as a function of threshold.

Two algorithms are provided to minimise the penalised negative log likelihood in Equation~\ref{eq:HTPNLL}. The default is a Newton-Raphson approach exploiting gradients; the alternative is a simplex search procedure.

Once parameter estimates are available, we represent the distribution of the $(D-1)$-dimensional residual random variable $\un{Z}$ using the set $\mathcal{E}_b=\{e_{i'bd'}\}$, for all $d'=2,3,...,D$, and all $i'=1,2,...,N$ such that $y_{i'1}>\phi(\tilde{\tau})$, $b=A(i')$, of residuals from the fit per covariate bin, with elements
\begin{eqnarray}
	e_{i'bd'} = \left( y_{i'd'} - \hat{\alpha}_{bd'} y_{i'1} - \hat{\mu}_{d'} y_{i'1}^{\hat{\beta}_{d'}} \right) /
	\left( \hat{\sigma}_{d'} y_{i'1}^{\hat{\beta}_{d'}} \right) .
\end{eqnarray}
During subsequent simulation under the fitted conditional extremes model, these residuals are resampled jointly as $\{e_{i'b2}, e_{i'b3}, ..., e_{i'bD}\}$, thereby preserving the dependence between them, and hence the dependence between variables $Y_2, Y_3, ..., Y_D|(Y_1=y$, $y>\phi(\tilde{\tau}), \mathcal{B}=b)$ per covariate bin. Similarly, simulation below threshold $\phi(\tilde{\tau})$ is achieved simply by resampling from the original Laplace-scale storm peak sample. We note that \PPC also facilitates estimation of variants of this dependence model, for which any number of conditional extremes model parameters $\alpha$, $\beta$, $\mu$ and $\sigma$ are allowed to vary between covariate bins, and their overall roughness penalised for good performance by extending the penalty term in Equation~\ref{eq:HTPNLL} to include all ``non-stationary'' parameters; see Section~4.2 of the \PPC user guide. It is also possible to pool estimates of residuals across covariate bins, useful when covariate bins with low occupancy are present.

\textbf{Uncertainty quantification} for extremal dependence inference is performed by extending the bootstrap procedure described in Section~\ref{Sct:OvrSft:MrgMdl}. Conditional extremes models are estimated for each of the bootstrap resamples based on which marginal models were estimated. The combined marginal and conditional extremes analysis therefore produces the set $\{ \{ \xi_d^r, \{\nu_{bd}^r, \psi_{bd}^r, \omega_{bd}^r, \kappa_{bd}^r, l_{bd}^r\}_{b=1}^B\}_{d=1}^D, \{ \{\alpha_{bd'}^r\}_{b=1}^B, \beta_{d'}^r, \mu_{d'}^r, \sigma_{d'}^r,  \}_{d'=2}^D, \{\mathcal{E}_{b}^r\}_{b=1}^B; \{\tau_d^r, \lambda_d^\circ \}_{d=1}^D, \{\tilde{\tau}_{d'}^r, \tilde{\lambda}_{d'}^\circ \}_{d'=2}^D \}$ of parameter estimates and residuals for each of $R$ bootstrap resamples indexed by superscript $r$. As for marginal modelling, the value of roughness coefficient  $\tilde{\lambda}_{d'}^\circ$ is typically estimated using the original sample only, and adopted for all bootstrap resamples also, although the option to estimate a new optimal roughness coefficient per bootstrap resample is provided.

A sampling procedure is used to estimate the \textbf{conditional return value} distribution of the associated value for $\dot{Y}_{d'}$ given a $T$-year maximum event for $\dot{Y}_1$ per covariate bin, and over unions of covariate bins. The approach combines importance sampling from the marginal $T$-year maximum distribution of $\dot{Y}_1$ (exploiting Equation~\ref{eq:MrgTYer}) with Monte Carlo sampling under the fitted conditional extremes model, managing transformations between marginal and standard Laplace scales. Further discussion of conditional return values can be found in \cite{TowEA23}.

\subsection{Estimation of design contours} \label{Sct:OvrSft:EstDsgCnt}
Using the estimated marginal and conditional extremes models, \PPC allows estimation of quantities such as return values for the dominant variable, conditional return values for associated variables, and environmental design contours. \cite{HslEA17}, \cite{RssEA19} and \cite{HslEA21} provide recent reviews of contour estimation, encompassing a range of approaches. Algorithms to estimate three types of design contour are implemented in \PPC. These are (a) constant exceedance contours; (b) direct sampling contours; and (c) conditional extremes (or ``HT'', in acknowledgement of the authors of the conditional extremes model, Heffernan and Tawn) constant density contours. Each of the three contour methods estimates a line (in 2-D) or surface (in 3-D) on which certain characteristics of the distributions of $\dot{Y}_2|(\dot{Y}_1=y)$ (in 2-D, or $(\dot{Y}_2, \dot{Y}_3)|(\dot{Y}_1=y)$ in 3-D) for large $y$ are preserved. For example, the constant exceedance contour in 2-D is a line on which the probability of exceedance in an ``outward'' sense (e.g. $\mathbb{P}(\dot{Y}_1>\zeta_1,\dot{Y}_2>\zeta_2)$ in the first quadrant) is constant, for points $(\zeta_1, \zeta_2)$ on the contour; see \cite{JntEwnFln12b} for details. The direct sampling contour of \cite{HsbEA15a} is similar, except that now the probability in the half plane defined by the tangent to the contour at any point of interest is constant, and the contour itself must be convex. In 2-D, the conditional extremes constant density contour defines a line on which the joint density of the pair $(\dot{Y}_1, \dot{Y}_2)$ is constant. By construction, all contours pass through a so-called ``lock point'', defined as an extreme quantile (e.g. the $T$-year return value) of dominant variable $Y_1$ and the corresponding conditional median of associated variate $Y_2|$($Y_1=T$-year value). The lock point defines the value of the distributional characteristic preserved on the contour. Further discussion and illustrations are provided in Section 5 of the \PPC user guide. {The contour methods discussed in this section are intended to illustrate the kinds of inferences which can be made routinely using simulation under a fitted \PPC model. All the approaches discussed provide means to quantify the extent of some or all of a cloud of realisations. No claims are made regarding their usefulness in any particular field of application (e.g. \citealt{MckHsl21}, \citealt{HfvEA22}). The user interested in making specific inferences for particular structure variables, or choices of design contour, is free to implement these as alternatives or extensions to the existing Stage 5 of the analysis procedure}

Mathematically, in 2-D, the constant exceedance contour $\un{\zeta}(\theta) = (\zeta_1(\theta),\zeta_2(\theta))$ for $\theta \in \Theta \subseteq [0,360)$ can be defined by the equation
\begin{eqnarray}
	\mathbb{P}\left(\bigcap_{d=1}^2 \left(r_d(\theta) \dot{Y}_d>r_d(\theta) \zeta_d(\theta)\right)\right) = p
\end{eqnarray}
where $\un{r}^\circ = (r_1^\circ, r_2^\circ)$ is a reference location (see e.g. \citealt{JntEwnFln12b}), and $r_d(\theta) = \zeta_d(\theta)-r_d^\circ$, $d=1,2$, for some small probability $p>0$. {We set the value of $r_2^o$ equal to the conditional mean $\dot{Y}_2|(\dot{Y}_1=$T-year value$)$, and the value of $r_1^o$ to the minimum value of $\dot{Y}_1$ in the sample, to ensure continuity of the $T$-year contour when $\dot{Y}_1$ exceeds the sample minimum.} 

{To estimate the direct sampling contour in two-dimensions for probability level $\alpha$, following \cite{HsbEA15a}, we first find the function $C(\theta)$, the $(1-\alpha)$-quantile of the distribution of the projection $\dot{Y}_1 \cos(\theta)+\dot{Y}_2 \sin(\theta)$ for each value of $\theta\in[0,2\pi)$
	\begin{eqnarray}
		C(\theta)=\inf\left\{c:\mathbb{P}\left(\left[\dot{Y}_1 \cos(\theta)+\dot{Y}_2 \sin(\theta)\right]>c\right)=\alpha\right\} .
	\end{eqnarray}
	Then we estimate the contour $\mathcal{C}=\{(\zeta_1(\theta),\zeta_2(\theta)) : \theta \in [0,2\pi)\}$ using
	\begin{eqnarray}
		\zeta_1(\theta) &=& C(\theta) \cos(\theta) - \frac{dC}{d\theta}\sin(\theta) , \\
		\zeta_2(\theta) &=& C(\theta) \sin(\theta) + \frac{dC}{d\theta}\cos(\theta) ,
	\end{eqnarray}
	and potentially further smooth $\mathcal{C}$ as a function of $\theta$. Following \cite{WntEA93}, for at $T$-year return period, it is recommended that the value of $\alpha$ be set to $1/T$.}

The conditional extremes constant density contour consists of the set of points $\un{\zeta} \in \mathbb{R}^2$ for which the joint density $f_{\un{\dot{Y}}}(\un{\zeta}) = c$ for some small value $c>0$. Algorithms for estimation of the three contour types exploit importance sampling (e.g. Section 3.4 of \citealt{TowEA21}) for computational efficiency where possible. Contours can be estimated per covariate bin (e.g. as ``directional'' or ``seasonal'' contours), or integrated over covariate bins (e.g. to provide ``omni'' estimates over all bins). 

{The direct sampling contour is convex by definition, in the sense that it encloses a convex region (of $\mathbb{R}^2$  for 2-D contours), possibly with the help of the coordinate axes}. Depending on the nature of the joint density, the conditional extremes constant density ``contour'' could exist in the form of a set of disjoint curves, some of which may be closed. The constant exceedance and direct sampling contours are invariant to transformations of variables, whereas the conditional extremes constant density contour is not (e.g. \citealt{RssEA19}).

\subsection{Accessing the software, and performing a typical analysis} \label{Sct:OvrSft:PrfTpcAnl}
The \PPC is available for download from GitHub at \cite{TowEA23a}. Software is written using MATLAB object-oriented programming. Full specifications of classes, properties and methods are therefore provided, and available for user inspection.  

A typical analysis involves executing each of the MATLAB scripts \texttt{Stage1\_PeakPicking}, \texttt{Stage2\_SetBinEdges}, \texttt{Stage3\_FitMargin} (for each variable marginally), \texttt{Stage4\_FitHeffernanTawn} and \texttt{Stage5\_Contour} in sequence, as illustrated by the workflow in Figure~\ref{Fgr:covXtremeFlowScheme}. As explained in detail in the \PPC user guide, each stage of the analysis involves the specification of control parameters for that stage. Default values for control parameters are provided, but it is necessary for the user at each stage to assess whether values are appropriate by confirming that diagnostic plots generated have reasonable characteristics. It may be necessary to repeat a given stage multiple times until acceptable diagnostic characteristics are found, before proceeding to the next stage.

Sections~\ref{Sct:CasStd1} and \ref{Sct:CasStd2} below provide brief descriptions of the application of \PPC to a pair of random variables (specifically significant wave height and spectral peak period) varying with directional covariate (Section~\ref{Sct:CasStd1}) and to four random variables (significant wave height, spectral peak period, wind speed and overturning momemt) varying with 2-D directional-seasonal covariate.

\FloatBarrier
\section{Case study : Single covariate, bivariate response} \label{Sct:CasStd1}
This section outlines to application of \PPC to estimation of design contours for extreme storm peak significant wave height and associated spectral peak period, both varying with storm direction. The analysis follows the five stages discussed in Section~\ref{Sct:OvrSft}. The data correspond to approximately 35 years of time-series output from a hindcast simulator for a location in the northern North Sea.

\begin{figure}[h!]
	\centering
	\includegraphics[width=0.9\textwidth]{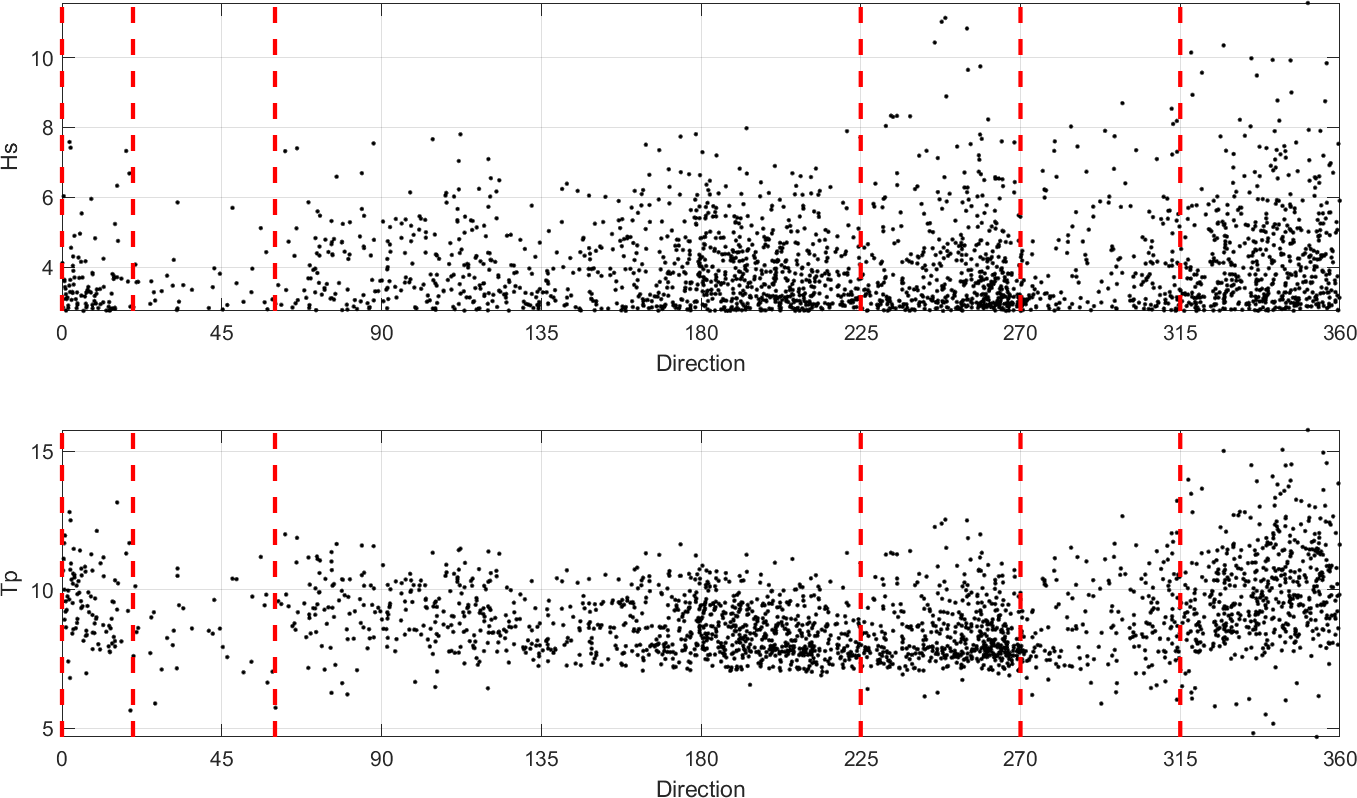}
	\caption{Directional variation of storm peak significant wave height ($H_S$ in metres, top) and associated spectral peak period ($T_P$ in seconds, bottom). {Directions (in degrees) indicate the angle from which a storm emanates, defined clockwise from North}. Also shown are directional bin edges (red) for 6 bins. The variation of $H_S$ and $T_P$ is approximately independent of direction within each bin, but different between bins.}
	\label{Fgr:CS1:DrcPlt}
\end{figure}

\begin{figure}[h!]
	\centering
	\includegraphics[width=0.9\textwidth]{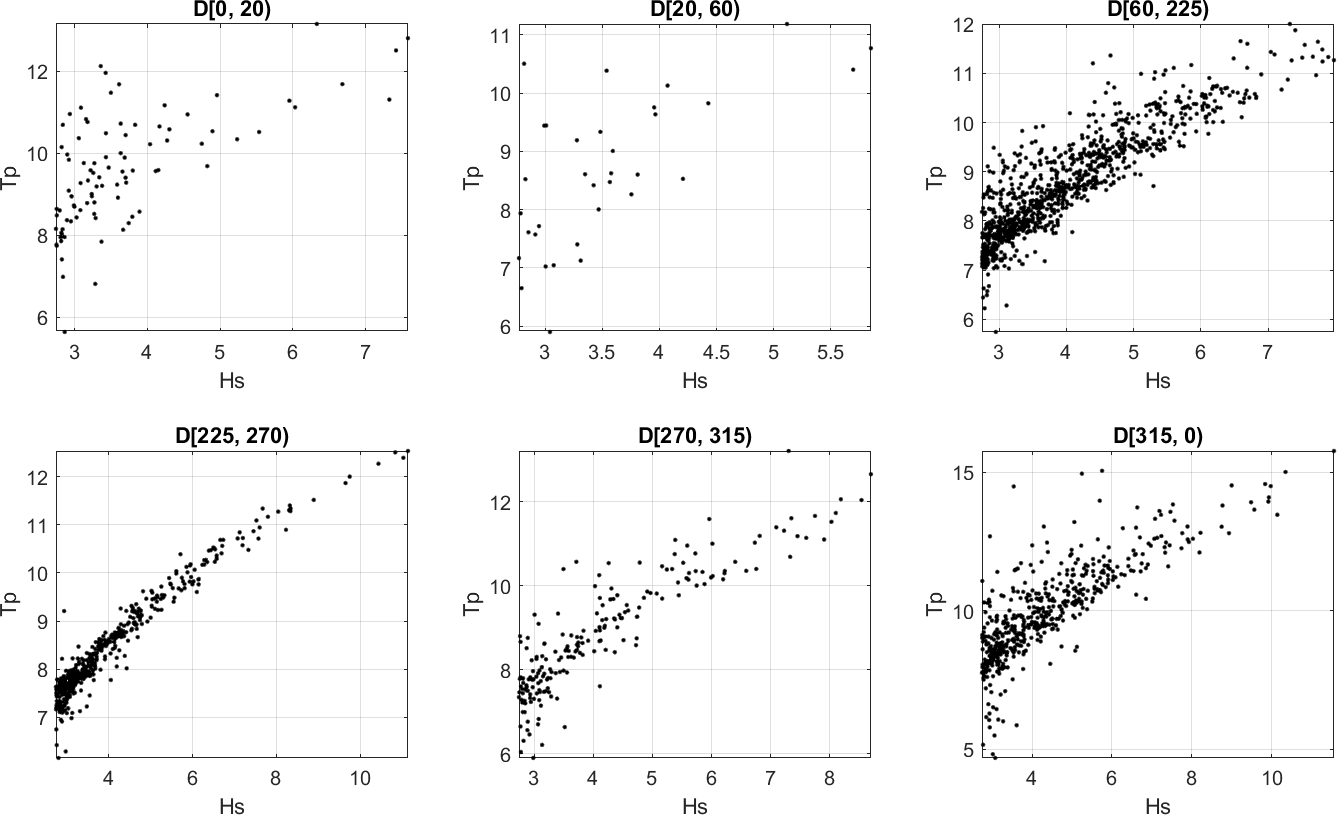}
	\caption{Associated spectral peak period (s) on storm peak significant wave height (m) per directional bin. Panel titles indicate that the covariate is direction ``D'', and give the angular interval corresponding to the bin. It is apparent that the dependence between $H_S$ and $T_P$ varies between bins.}
	\label{Fgr:CS1:DrcSctPlt}
\end{figure}

Following isolation of storm peak significant wave height $H_S$, associated spectral peak period $T_P$ and storm peak direction using Stage 1, Figure~\ref{Fgr:CS1:DrcPlt} illustrates the output of Stage~2 of the analysis, showing storm peak $H_S$ and associated $T_P$ as a function of the direction from which storms emanate, in degrees clockwise from north. The figure also shows the user-input bin edges at $0^\circ$, $20^\circ$, $60^\circ$, $225^\circ$, $270^\circ$ and $315^\circ$, specified so that the variation of $H_S$ and $T_P$ is approximately independent of direction within each bin, but different between bins. Thus, for example, the bin including $45^\circ$ corresponds to the land shadow of Norway, with resulting low values of storm peak $H_S$, whereas the bins including $250^\circ$ and $340^\circ$ contain large storms from the Atlantic Ocean and Norwegian Sea.
Figure~\ref{Fgr:CS1:DrcSctPlt} provides scatter plots of associated $T_P$ on storm peak $H_S$ for each of the six directional bins. The rates of occurrence of storms, and the marginal characteristics of $H_S$ and $T_P$ are clearly different between bins. In directional bin $[225,270)$, the dependence between $T_P$ and $H_S$ appears to be particularly strong. 

\begin{figure}[h!]
	\centering
	\includegraphics[width=0.9\textwidth]{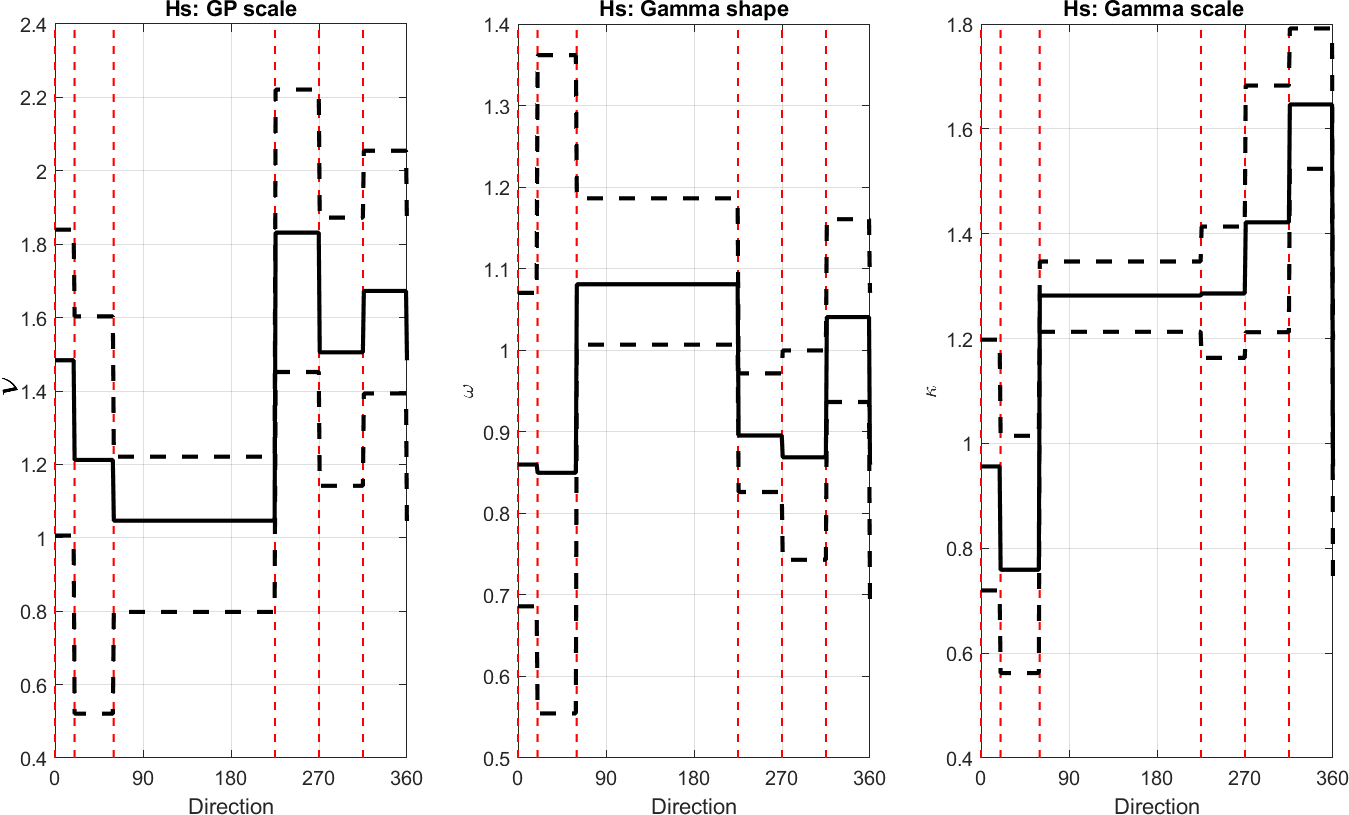}
	\caption{Marginal directional extreme value model for storm peak significant wave height (m). Variation of parameter estimates for GP scale $\nu$, gamma shape $\omega$ and scale $\kappa$ with direction. Mean estimates in solid black, and bootstrap 95\% uncertainty bands in dashed black. Also shown are directional bin edges (red). Directional dependence in particularly clear for $\nu$ and $\kappa$. {See Equations~\ref{eq:DnsGmm} and \ref{eq:DnsGP} for model forms and parameter definition.}}
	\label{Fgr:CS1:HsByDrcPrm}
\end{figure}

\begin{figure}[h!]
	\centering
	\includegraphics[width=0.9\textwidth]{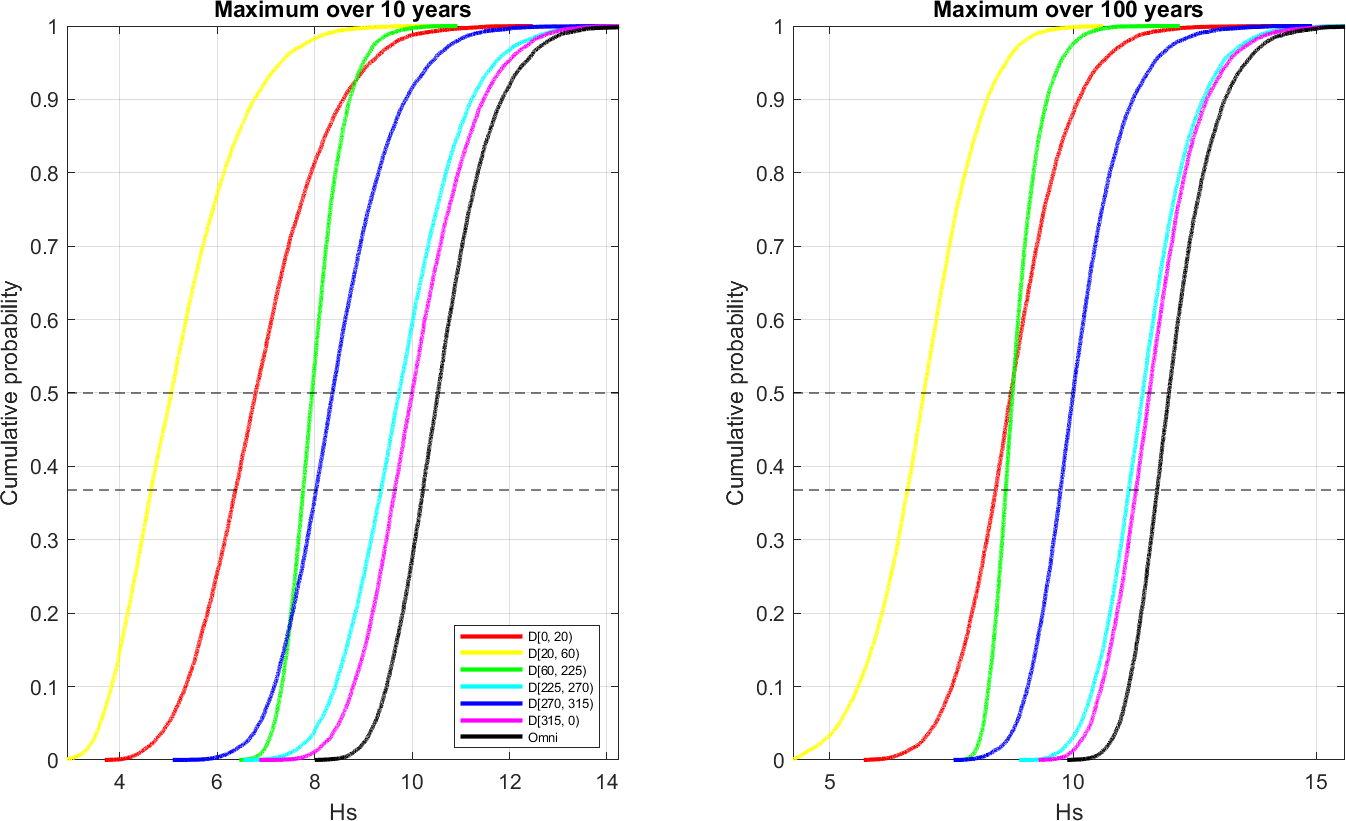}
	\caption{Cumulative distribution functions for the 10-year (left) and 100-year (right) maximum of storm peak significant wave height (m) per directional bin and over all bins (``omni'', black). Horizontal dashed lines drawn at the $\exp(-1)$ quantile and median.}
	\label{Fgr:CS1:HsRtrVls}
\end{figure}

Figure~\ref{Fgr:CS1:HsByDrcPrm} illustrates marginal extreme value models for storm peak $H_S$, the panels describing the variation of the estimates of generalised Pareto scale $\nu$, gamma shape $\omega$ and scale $\kappa$ with direction, in terms of means (solid) and 95\% uncertainty bands (dashed) over bootstrap resamples and random choices of marginal threshold non-exceedance probabilities drawn from the interval $[0.7,0.9]$. The empirical density of corresponding (stationary) estimates of generalised Pareto shape $\xi$ is shown in Figure~9 of the user guide to be Gaussian-like with mean at approximately -0.2. The variation of $\nu$ and $\kappa$ with direction appears consistent with expectations given Figures~\ref{Fgr:CS1:DrcPlt} and \ref{Fgr:CS1:DrcSctPlt}. In particular, given estimated $\nu$, the largest return values for storm peak $H_S$ would be expected to emanate from the Atlantic and Norwegian Sea. Indeed, inspection of directional maxima for storm peak $H_S$ for return periods of 10 and 100 years in Figure~\ref{Fgr:CS1:HsRtrVls} confirms this: the largest contributors to the omni-directional maximum (in black) are the covariate bins corresponding to the Norwegian Sea ($[315,0)$, magenta) and Atlantic ($[225,270)$, cyan).

It is critical to assess the diagnostic plots generated by \PPC to confirm that model fit is adequate. A number of illustrative diagnostic plots corresponding to this case study are shown in the user guide. For the current application, a plot of the predictive negative log likelihood from the cross-validation procedure (see Equation~\ref{eq:LamMM}) suggests that the optimal choice $\lambda^\circ$ of roughness penalty lies at around three. The goodness of fit of the marginal model is assessed by examining the stability of the estimate for generalised Pareto shape parameter $\xi$ as a function of extreme value threshold $\tau$. Comparison of empirical tails directly from the sample, with corresponding tails (and their uncertainty) estimates under the extreme value model, per covariate bin and omni-directionally, also suggests the marginal model is reasonable. The corresponding full marginal analysis must also be performed for associated $T_P$ with direction. 

\begin{figure}[h!]
	\centering
	\includegraphics[width=0.9\textwidth]{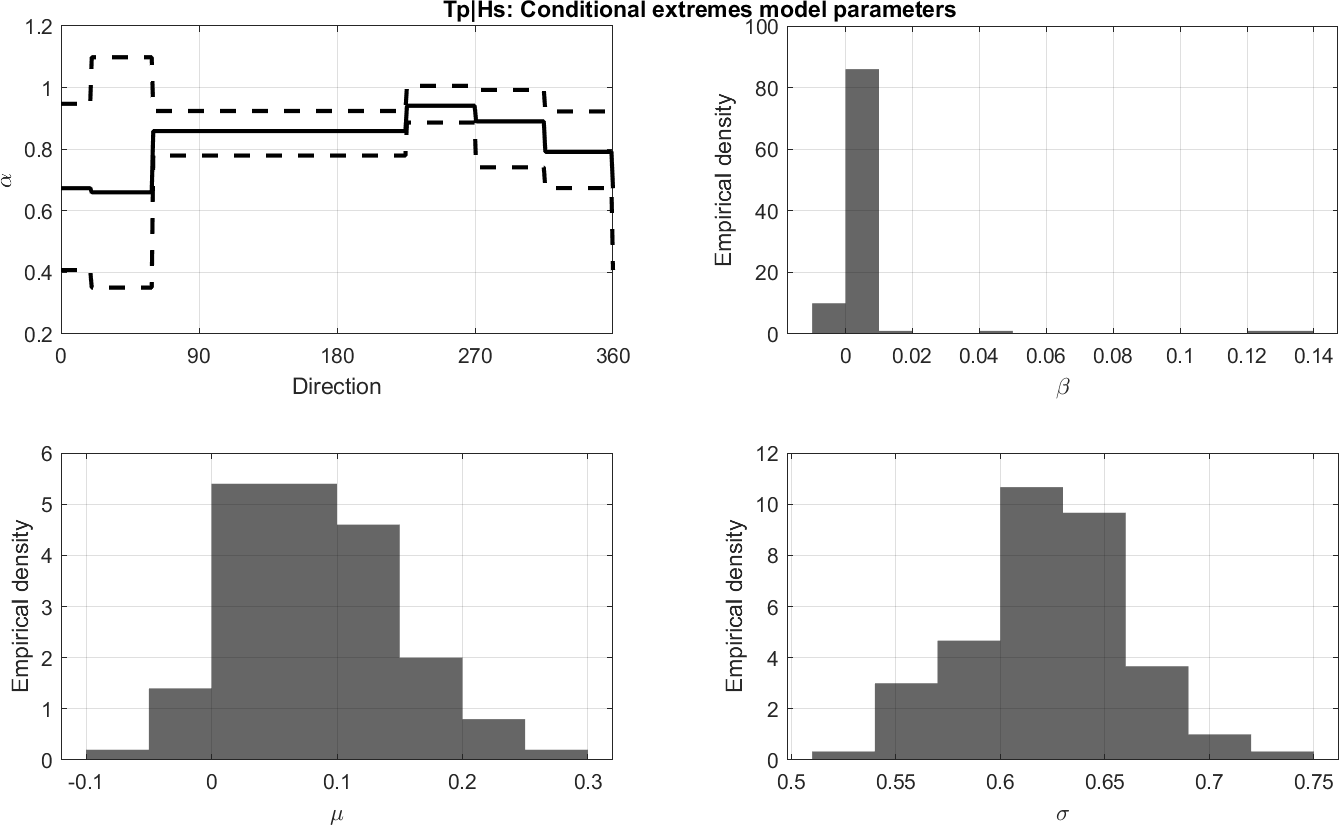}
	\caption{Parameter estimates from the conditional extreme value mode for associated peak period given storm peak significant wave height (m). {Top left: directional variation of estimated linear slope parameter $\alpha$ summarised as mean and 95\% bootstrap uncertainty band; top right: histogram of estimated exponent parameter $\beta$; bottom left: histogram of estimated residual mean $\mu$; bottom right: histogram of estimated residual standard deviation $\sigma$. See Equations~\ref{eq:HTMdl} and \ref{eq:HTRsd} for model form and parameter definition.}}
	\label{Fgr:CS1:HTPrm}
\end{figure}

Using the marginal models, Figure~\ref{Fgr:CS1:HTPrm} illustrates parameter estimates for the conditional extremes model for associated $T_P$ given storm peak $H_S$ on standard Laplace margins. The non-stationary estimate for slope parameter $\alpha$ suggests that dependence between the variables is high (near the maximum possible of unity), especially in the covariate bin corresponding to Atlantic storms. {The uncertainty band for the slope parameter included unity, admissible at sub-asymptotic levels (\citealt{TndEA21})}. The value of exponent parameter $\beta$ is slightly larger than zero, indicating that the sizes of residuals ``$y^\beta Z$'' from the conditional extremes model fit (Equation~\ref{eq:HTMdl}) grow very slowly with the conditioning value $y$ on Laplace scale. The empirical densities for residual parameters $\mu$ and $\sigma$ are typical in our experience; unimodal densities, with approximately Gaussian shape.

Diagnostic plots are again assessed to confirm reasonable goodness of fit for the conditional extremes model. For the current application, parameter estimates from $\alpha$ per covariate bin are reasonably stable as a function of dependence threshold $\tilde{\tau}$ on the interval $[0.7,0.85]$ (see Figure 20 of the user guide). Further, the distribution of residuals $\mathcal{E}$ from the model fit do not appear to be obviously dependent on the directional covariate (see Figure 18 of the user guide). The overall distribution of residuals from the model fit (see Figure 17 of the user guide) is typical; a Gaussian-like density with positive skew. 

\begin{figure}[h!]
	\centering
	\includegraphics[width=0.9\textwidth]{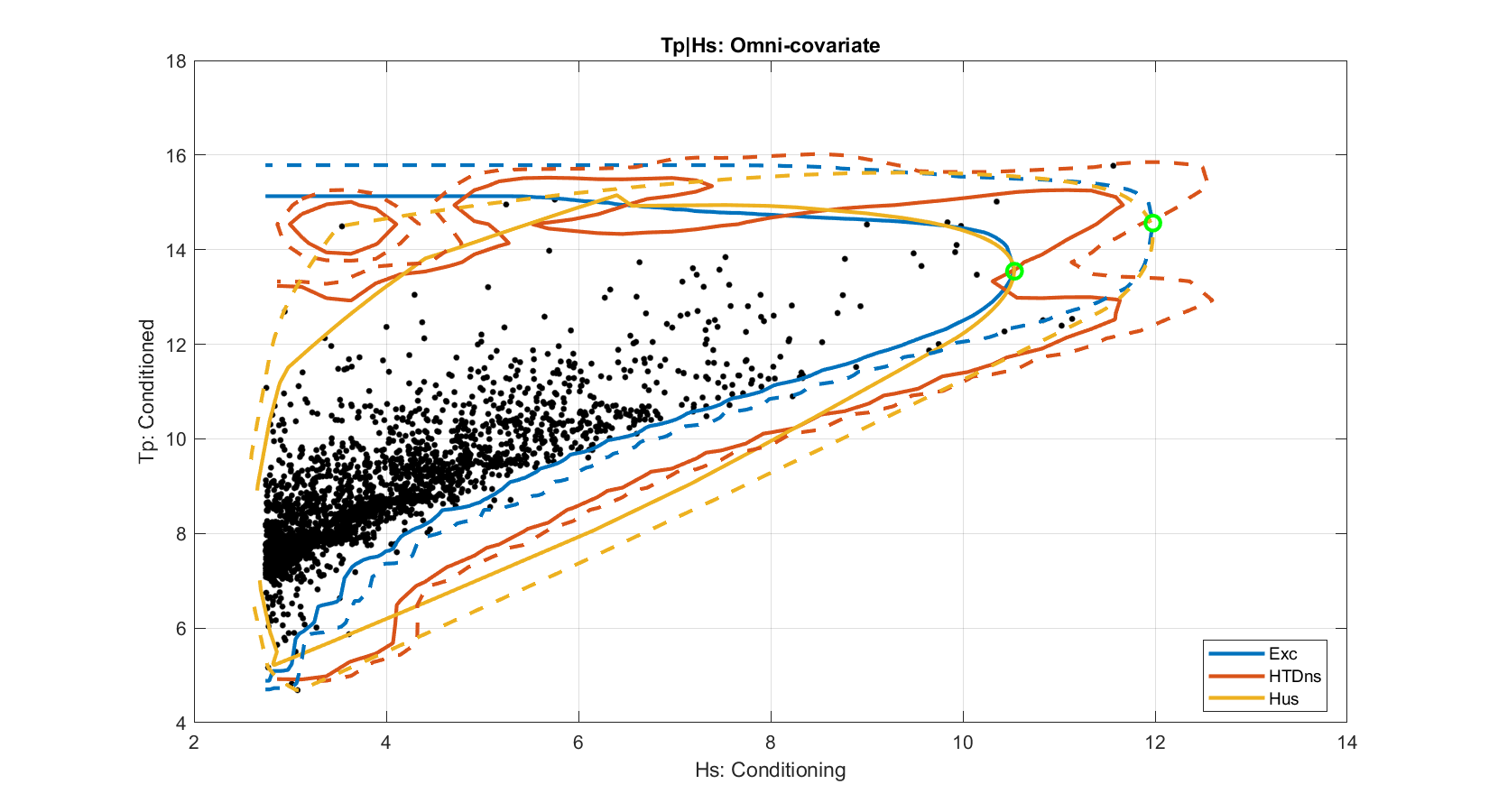}
	\caption{Omni-directional environmental contours of associated peak period (s) and storm peak significant wave height (m), for 10- and 100-year maximum values of storm peak significant wave height shown as solid and dashed lines respectively, corresponding to the Exceedance (Exc, blue), Heffernan-Tawn density (HTDns, orange) and Huseby (Hus, yellow) contours. Lock points for the respective return periods are shown in green.}
	\label{Fgr:CntOmni}
\end{figure}

\begin{figure}[h!]
	\centering
	\includegraphics[width=0.9\textwidth]{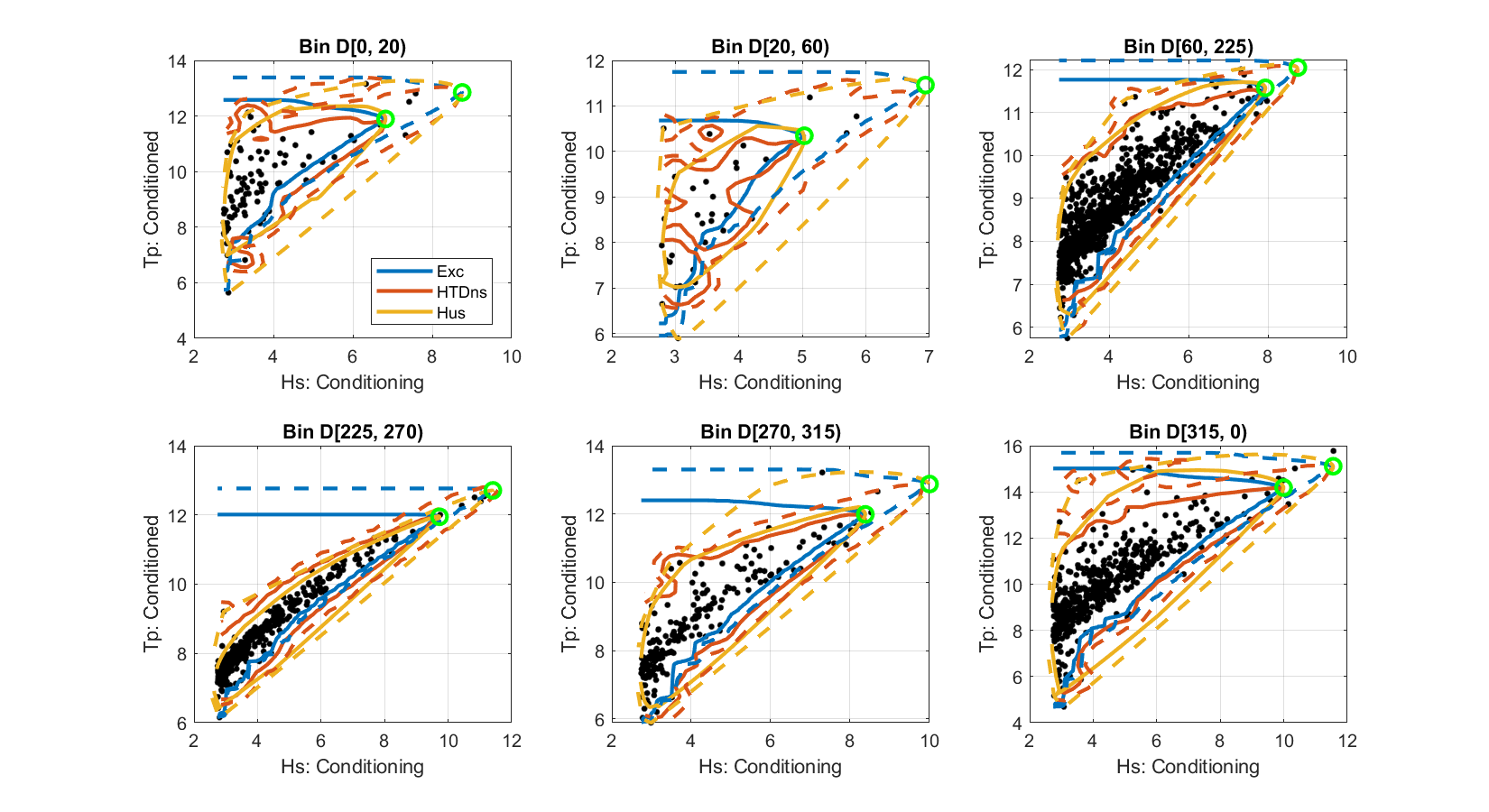}
	\caption{Environmental contours of associated peak period (s) and storm peak significant wave height (m), for individual directional bins. Contours shows for 10- and 100-year maximum values of storm peak significant wave height, shown as solid and dashed lines respectively, corresponding to the Exceedance (Exc, blue), Heffernan-Tawn density (HTDns, orange) and Huseby (Hus, yellow) contours. Lock points for the respective return periods are shown in green.}
	\label{Fgr:CntPerBin}
\end{figure}

We simulate under the fitted models to generate the environmental design contours (see Section~\ref{Sct:OvrSft:EstDsgCnt}) shown in Figure~\ref{Fgr:CntOmni} omni-directionally and Figure~\ref{Fgr:CntPerBin} per directional bin. The constant exceedance, direct sampling and HT density contours are labelled as ``Exc'', ``Hus'' and ``HTDns'' in the figures. The three different contour methods produce estimates which have similar characteristics in terms of describing the ``extent'' of the data cloud, reflecting the positive dependence between $H_S$ and $T_P$, and passing through the appropriate lock points (shown in green in the figures). However the methods also use different definitions of the environmental contour; it is not surprising therefore that the contours estimates do not agree fully. The \texttt{HTDns} contour in particular produces a more variable estimate, especially when sample size is small (e.g. in the [20,60) directional bin in Figure~\ref{Fgr:CntPerBin}). For engineering design, points on the contour with large values of $H_S$ would typically be adopted to test the integrity of models for the offshore or coastal structure of interest.

\FloatBarrier
\section{Case study : 2-D covariate, multivariate response}  \label{Sct:CasStd2}
The second case study is an extension of that reported in Section~\ref{Sct:CasStd1}, which incorporates the over-turning moment (OTM) experienced by an offshore structure subjected to wind and wave loading. Specifically, we seek to characterise the joint distribution of (storm peak) $H_S$ and wind speed (WS) conditional on extreme values of OTM, subject to directional and seasonal covariate effects. The analysis again follows the 5 stages described in Section~\ref{Sct:OvrSft}. 

\begin{figure}[h!]
	\centering
	\includegraphics[width=0.9\textwidth]{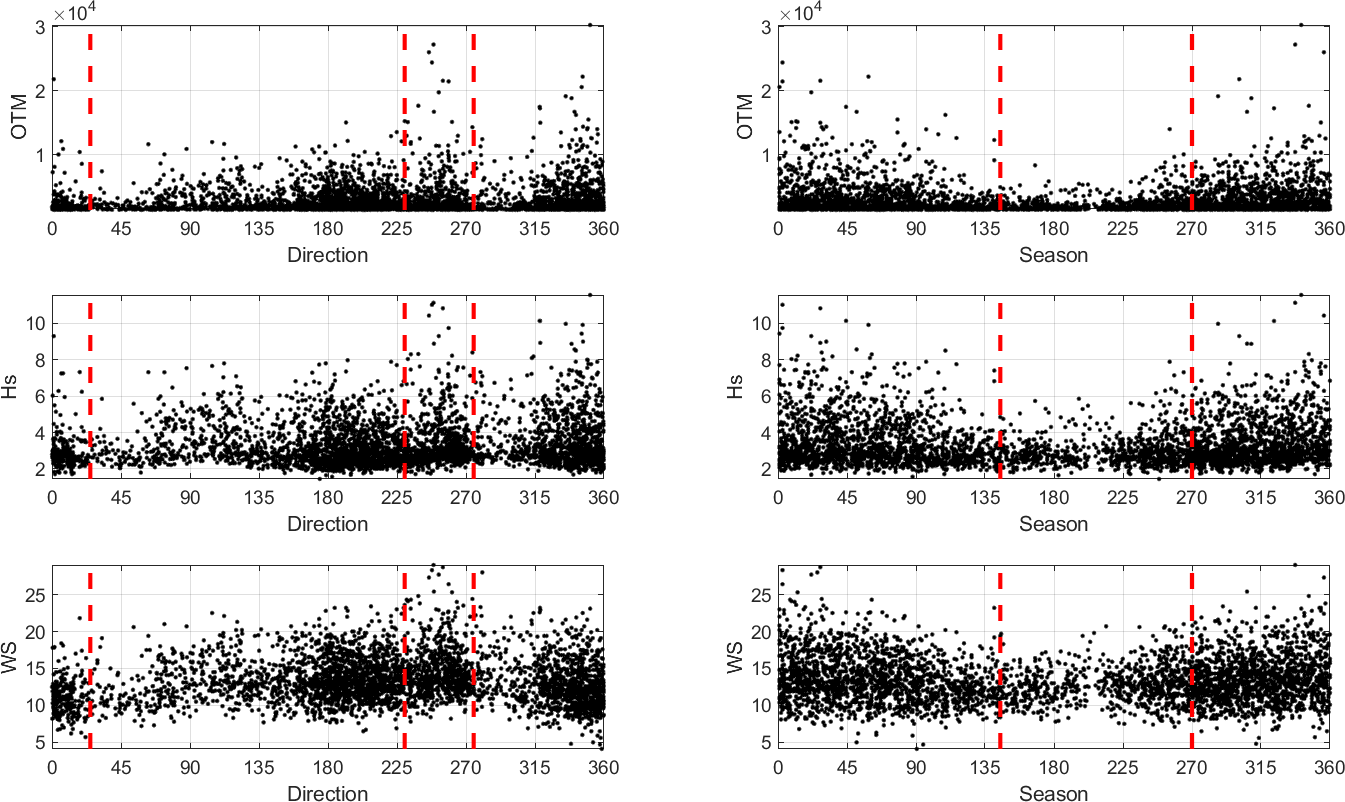}
	\caption{Directional (left) and seasonal (right) variation of overturning moment (OTM, in mega Newton metres, MNm, top), storm peak significant wave height ($H_S$ in m, second row) and associated wind speed (WS in metres per second, $\text{ms}^{-1}$, bottom). Also shown are bin edges for three directional bins coupled with two seasonal bins (and hence a total of 6 $=4 \times 2$ directional-seasonal bins). The variation of OTM, $H_S$ and WS is approximately independent of covariates within bins, but different between bins. For the directional convention, see caption of Figure~\ref{Fgr:CS1:DrcPlt}. {Season (or seasonal degree) is defined by mapping the calendar year linearly into (0,360], with 0 indicating midnight on the 1st January}.}
	\label{Fgr:CS2:CvrPlt}
\end{figure}

Stage 1 of the analysis is isolation of storm peak events from the underlying storm $H_S$ time-series data, following the same procedure as in Section~\ref{Sct:CasStd1}. For Stage 2, Figure~\ref{Fgr:CS2:CvrPlt} shows the directional and seasonal variation of the conditioning variate OTM, and associated variates $H_S$ and WS together with the directional and seasonal bin edges specified. Note that a smaller number of directional bins is used for this case study, to limit the total number of directional-seasonal bins and hence the number of parameters to be inferred in the analysis. Nevertheless, we are careful to allow for possible directional effects due to storms from the Atlantic and Norwegian Sea. There are clear directional and seasonal effects present.

\begin{figure}[h!]
	\centering
	\includegraphics[width=0.9\textwidth]{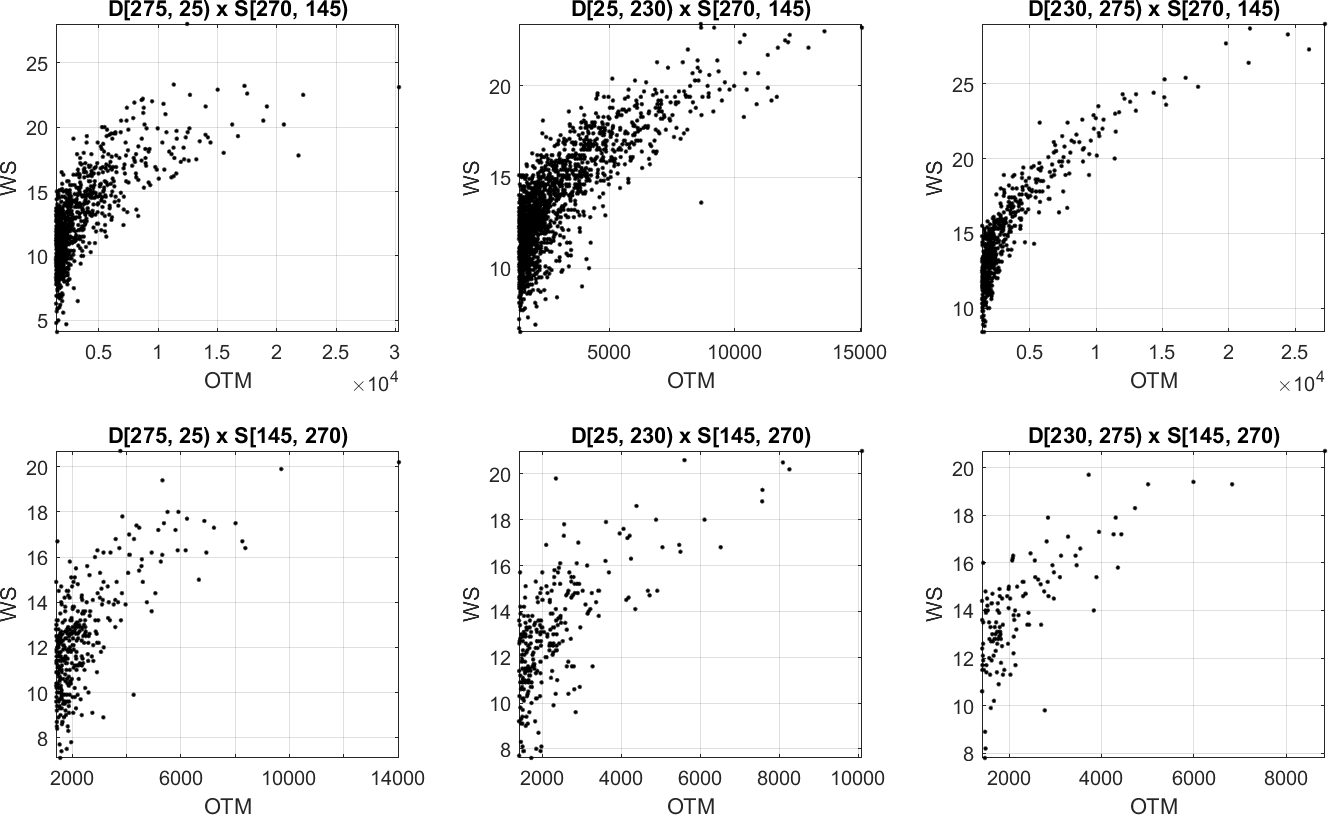}
	\caption{Associated wind speed ($\text{ms}^{-1}$) on storm peak overturning moment (MNm) per directional-seasonal bin. Panel titles indicate that the angular intervals of direction ``D'' and season ``S'' for the bin ``D''. It is apparent that the dependence between WS and OTM varies between bins, with obvious non-linearity in some bins. A similar plot is generated for associated significant wave height (m) on storm peak overturning moment.}
	\label{Fgr:CS2:CvrSctPltWSOTM}
\end{figure}

\begin{figure}[h!]
	\centering
	\includegraphics[width=0.9\textwidth]{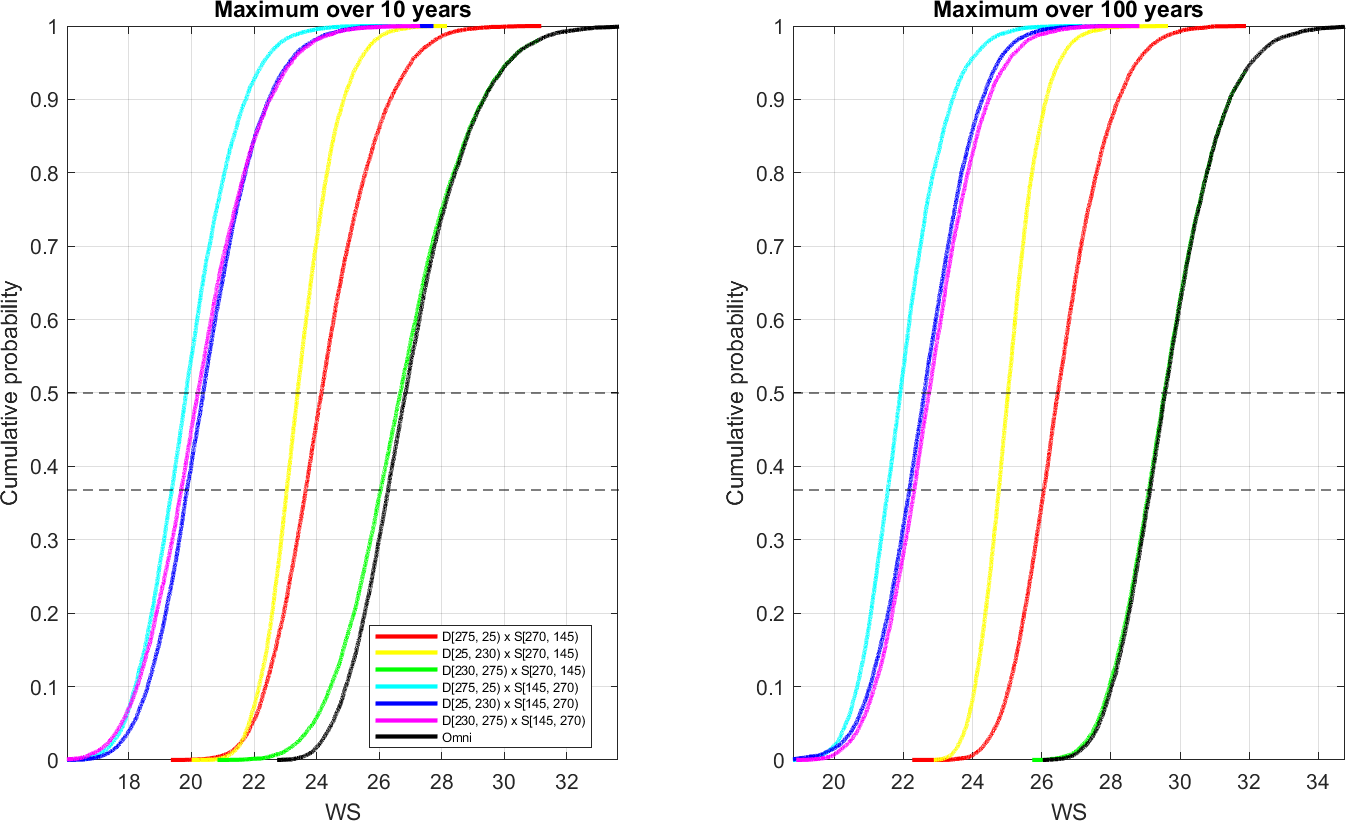}
	\caption{Cumulative distribution functions for the 10-year (left) and 100-year (right) maximum of associated wind speed ($\text{ms}^{-1}$) per covariate bin and over all bins (``omni'', black). Horizontal dashed lines drawn at the $\exp(-1)$ quantile and median.}
	\label{Fgr:CS2:WSRtrVls}
\end{figure}

The resulting scatter plots of WS on OTM per covariate bin are shown in Figure~\ref{Fgr:CS2:CvrSctPltWSOTM} (and the corresponding plot for $H_S$ on OTM in Figure 28 of the user guide). 
Illustrative diagnostic plots for the estimation of marginal models for each of $H_S$, OTM, and WS with direction and seasonal covariates are given in Figures 30-32 of the user guide. Marginal model parameters show clear directional and seasonal variation, and comparison of empirical and model-based tails suggest reasonable model fit. The fitted marginal models are then used to estimate the cumulative distribution functions for $T$-year maxima of interest; estimates for the 10- and 100-year maximum of WS are given in Figure~\ref{Fgr:CS2:WSRtrVls}. Unsurprisingly, the ``omni'' distribution (estimated over all directional and seasonal bins) is dominated by winter storms from the Atlantic sector, and the most probable 100-year maximum WS is approximately 29 ms$^{-1}$.

\begin{figure}[h!]
	\centering
	\includegraphics[width=0.9\textwidth]{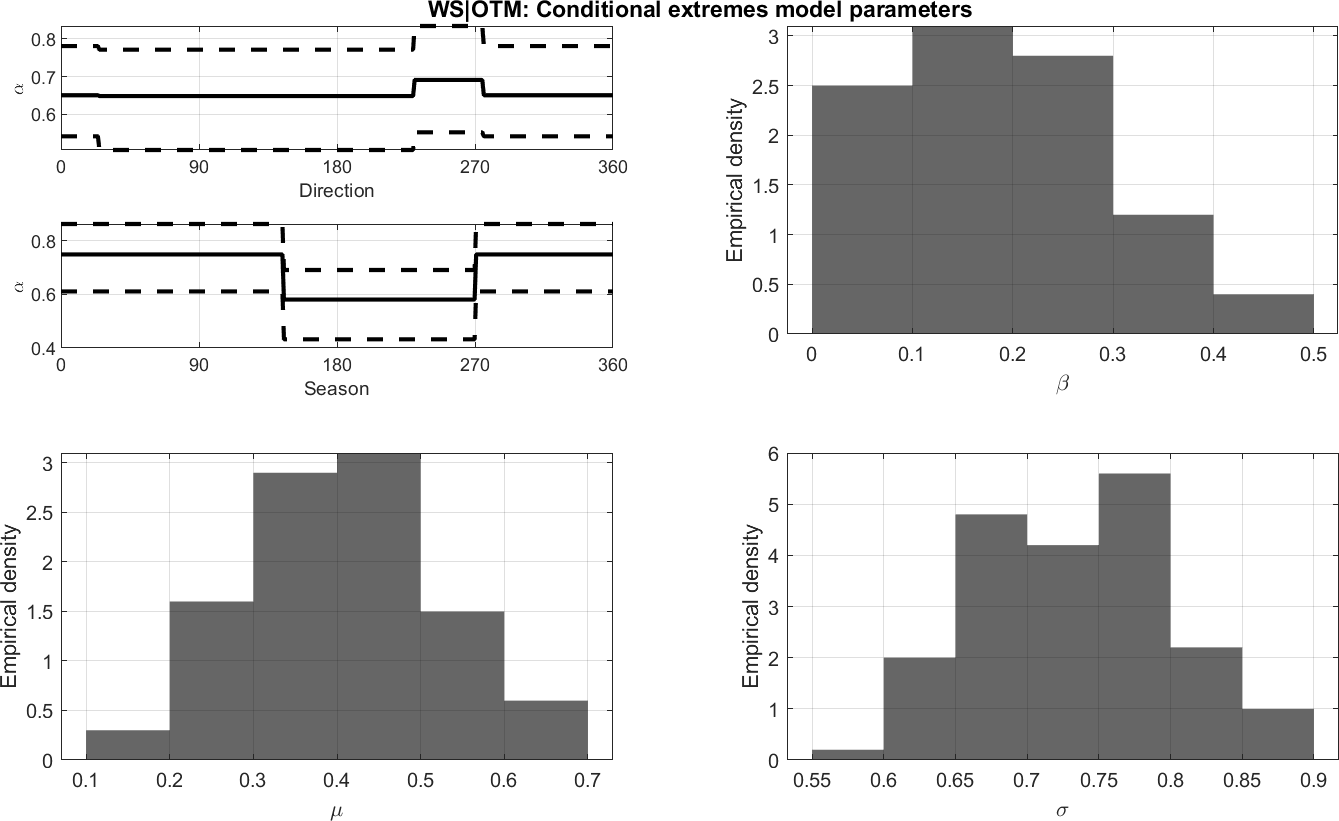}
	\caption{Parameter estimates from the conditional extreme value mode for associated wind speed given storm peak overturning moment. Top left: directional and seasonal variation of $\alpha$ summarised as mean and 95\% bootstrap uncertainty band; top right: histogram of $\beta$; bottom left: histogram of $\mu$; bottom right: histogram of $\sigma$. Note the very high directional dependence for $\alpha$ at around $280^\circ$. {See Equations~\ref{eq:HTMdl} and \ref{eq:HTRsd} for model form and parameter definition, and Figure~\ref{Fgr:CS1:HTPrm} for comparison.}}
	\label{Fgr:CS2:HTPrm}
\end{figure}

Following transformation to standard Laplace scale, a non-stationary conditional extremes model is estimated for $H_S$ and WS jointly, conditional on large OTM, given directional and seasonal covariates. Figure~\ref{Fgr:CS2:HTPrm} gives parameter estimates for WS, showing directional and seasonal dependence of the conditional extremes slope parameter $\alpha$; as would be expected perhaps, the dependence between WS and OTM is strongest for winter storms emanating from the North-West. The estimates for index $\beta$, common over covariate bins, is somewhat larger than found in Case Study 1. The empirical densities of nuisance parameters $\mu$ and $\sigma$ are approximately Gaussian-shaped. Supporting diagnostic plots for the conditional extremes model fit are given in Figures 33-38 of the user guide. In particular, Figure~37 indicates that the value of the $\alpha$ parameter is relatively stable for conditional extremes threshold levels $\tilde{\tau} \in [0.8,0.9]$, and Figure 38 indicates that a minimum roughness penalty for predictive lack of fit is achieved at around 10.

\begin{figure}[h!]
	\centering
	\includegraphics[width=0.9\textwidth]{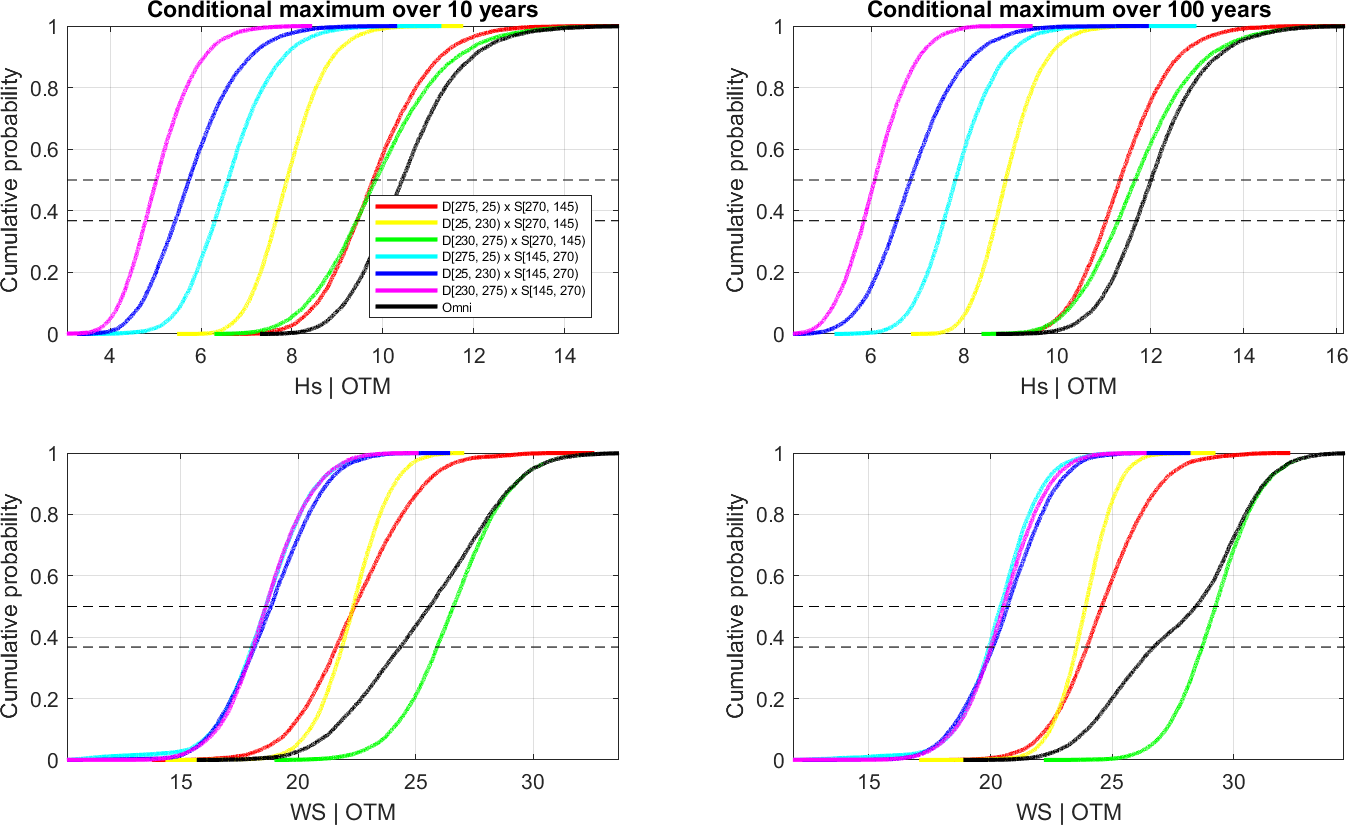}
	\caption{Conditional cumulative distribution functions of associated significant wave height (m, top) and associated wind speed ($\text{ms}^{-1}$ bottom) per directional bin and over all bins (``omni'', black), conditional on the 10-year (left) and 100-year (right) maximum of storm peak overturning moment (MNm). Horizontal dashed lines drawn at the $\exp(-1)$ quantile and median.}
	\label{Fgr:CS2:HsWSCndRtrVls}
\end{figure}

Figure~\ref{Fgr:CS2:HsWSCndRtrVls} gives estimated conditional return value distributions for $H_S$ and WS given 10-year and 100-year maximum OTM for individual directional-seasonal covariate bins, and ``omni'' over all covariate bins. For $H_S$, conditional return values are again largest for Atlantic winter storms. For WS however, we observe an interesting transition involving winter storms from directional sectors [275,315) to [230,275). The most probable \emph{conditional} value of WS given a 100-year maximum OTM is approximately 26.5 ms$^{-1}$, lower than the marginal most probable maximum 100-year wind speed.

\begin{figure}[h!]
	\centering
	\includegraphics[width=0.9\textwidth]{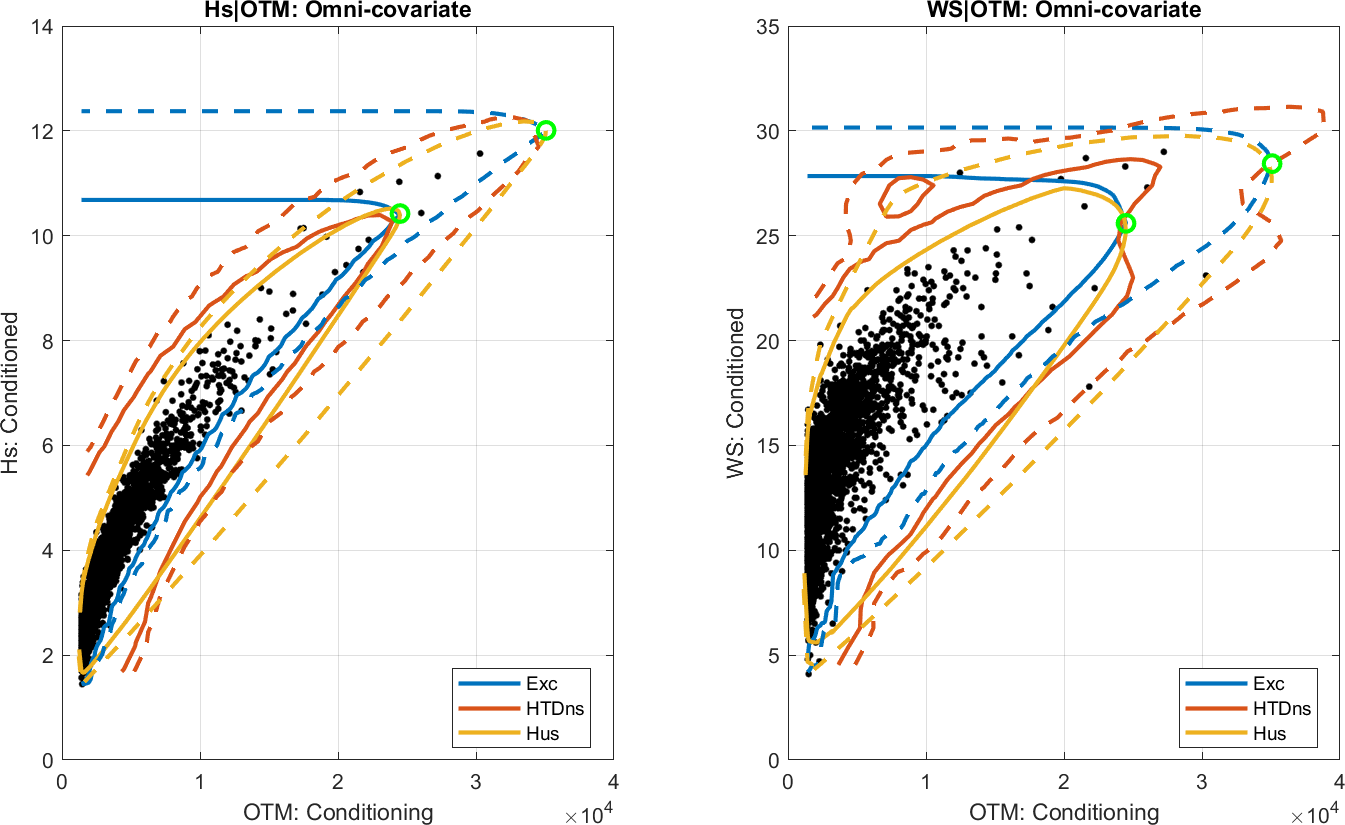}
	\caption{Omni-directional-seasonal environmental contours of associated significant wave height (m, left) and associated wind speed ($\text{ms}^{-1}$, right) and storm peak overturning moment (MNm). Contours shows for 10- and 100-year maximum values of storm peak significant wave height, shown as solid and dashed lines respectively, corresponding to the Exceedance (blue), Heffernan-Tawn density (HTDns, orange) and Huseby (yellow) contours. Corresponding contours per covariate bin are also generated.}
	\label{Fgr:CS2:HsWSCntOmni}
\end{figure}

The resulting omni-covariate environmental design contours for $H_S$ and WS, given occurrences of the 10-year and 100-year maximum OTM, are shown in Figure~\ref{Fgr:CS2:HsWSCntOmni}. The general features of the contours are similar to those of Figure~\ref{Fgr:CntOmni}. Corresponding illustrative plots of contours per covariate bin for WS are given in Figure 40 of the user guide.

\FloatBarrier
\section{Discussion} \label{Sct:Dsc}
This article introduces the \PPC software for pragmatic multivariate extreme value analysis with covariate non-stationarity. The MATLAB software provides functionality to isolate temporal peaks from time-series, and to define an appropriate partition of the covariate domain. Using this partition, marginal generalised Pareto models are estimated for each variate independently assuming piecewise constant threshold and scale parameterisations within a penalised likelihood framework; optimal scale parameter roughness is estimated using cross-validation. Marginal models are then used to transform the sample to standard Laplace margins. A non-stationary conditional extremes model is then estimated with piecewise constant parameterisation for the slope (``$\alpha$'') parameter, again using penalised likelihood estimation with cross-validation to estimate optimal roughness. Simulations and importance sampling are then used to estimate the distribution of $T$-year maxima for each variate, and environmental design contours conditional on a single conditioning variate. Uncertainty due to marginal and dependence threshold selection is quantified by fitting multiple models with randomly chosen thresholds within user-specified plausible intervals of threshold non-exceedance probability. Uncertainties in parameter estimates and subsequent inferences from model fitting are estimated using bootstrap resampling. As noted in Section~\ref{Sct:Int}, the software has already been used in a number of studies, mostly but not exclusively metocean-related.

The \PPC methodology makes a number of simplifying assumptions, motivated by the authors' experience of extreme value analysis applied to the ocean environment using a range of methodologies of difference complexities. For example, \PPC relies on sensible user-specified partitioning of the covariate domain into bins within which it is reasonable to assume common marginal tails and a common dependence structure; this simplifies inference considerably compared with competitor approaches. Moreover, we believe that inferences using \PPC with good partitioning are competitive with alternatives using more sophisticated tools. Specifically, \PPC marginally is equivalent to a Voronoi set representation with pre-specified covariate partition, which was demonstrated by \cite{ZnnEA19a} to be competitive with P-spline and Bayesian adaptive regression spline covariate representations. \PPC further assumes that the generalised Pareto shape parameter $\xi$ in each marginal model is constant with covariate; because of the relative difficulty of estimating the shape parameter compared with the scale, this would appear reasonable in the absence of strong evidence to the contrary, especially for small samples of data. Likewise, the $\beta$, $\mu$ and $\sigma$ parameters of the conditional extremes model are assumed stationary. This appears reasonable since $\beta$ is an exponent, again difficult to estimate. Moreover, $\mu$ and $\sigma$ are essentially nuisance parameters; any model misspecification caused by the assumption of stationarity will be accommodated to some extent by the adoption of residuals from model fitting for inferences under the model. It might be appropriate to relax some of the assumptions for specific applications, for example (a) when there is strong evidence that generalised Pareto $\xi$ is unlikely to be constant (e.g. due to land shadow and fetch limitation effects on $H_S$), or (b) since parameter estimates for conditional extremes $\alpha$ and $\mu$ are highly correlated when $\beta$ is close to unity. {We also note that it is not clear whether non-stationary marginal extreme analysis is necessarily the best approach to estimate marginal return values from non-stationary data. Provided that sufficient sample is available, so that a sufficiently high threshold for peaks over threshold analysis can be set, often a stationary marginal analysis is at least competitive if not preferable; see \cite{MckJnt20} for further discussion.}  

We anticipate that \PPC might provide a pragmatic starting point to studies of spatial and temporal dependence of extremes, since the underlying methodology of both conditional spatial extremes (e.g. \citealt{ShtEA19}, \citealt{ShtEA21}, \citealt{WdsTwn19}) and Markov extremal and similar time-series models (e.g. \citealt{WntTwn17}, \citealt{TndEA18}, \citealt{TndEA23}) involves each of Stages 1-4 of the \PPC approach. We also hope that \PPC might be useful generally in estimating joint characteristics of extremes from multivariate time-series.

\section*{Software and data availability}
\begin{itemize}
	\item Name: \PPC
	\vspace{-8pt}
	\item Developer: Ross Towe, Emma Ross, David Randell, Philip Jonathan
	\vspace{-8pt}
	\item Contact: ross.towe@shell.com
	\vspace{-8pt}
	\item First published: 2023
	\vspace{-8pt}
	\item Language: MATLAB (only base MATLAB needed)
	\vspace{-8pt}
	\item GitHub: \url{https://github.com/sede-open/covXtreme}
	\vspace{-8pt}
	\item Program size: 300kB  
	\vspace{-8pt}
	\item Licence: apache 2.0
	\vspace{-8pt}
	\item Free-to-use software, sample data and user guide available on GitHub.  
\end{itemize}

\section*{Credit}
Towe: Methodology, Software, Validation, Writing; Ross: Conceptualisation, Methodology, Software, Validation, Writing; Randell: Conceptualisation, Methodology, Software.
Jonathan: Conceptualisation, Methodology, Software, Validation, Writing.

\section*{Acknowledgement} \label{Sct:Dsc}
The original version of this software was developed as a component of the “Environmental Contours for SAfe DEsign of Ships and other marine structures” (ECSADES) project, part-funded by the European Union ERANET programme, and was summarised in a review paper on the definition and application of environmental contours (\citealt{RssEA19}). \PPC software, user guide and test data sets are available at \cite{TowEA23a}. {We would like to thank Jennifer Kensler at Shell for an internal technical review of \PPC methodology and software. At time of publication of this article, the software is hosted by Linux Foundation Energy at \url{https://lfenergy.org/projects/covxtreme/}, where the user community is encouraged to develop and share enhancements.}

\bibliographystyle{cas-model2-names}
\bibliography{C:/Users/Philip.Jonathan/PhilipGit/Code/LaTeX/phil}

\end{document}